\documentclass{aa}  

\usepackage{graphicx}

\usepackage{color}

\usepackage{pgf}
\usepackage{tikz}
\usepackage{pgfplots} 
\usepackage{txfonts}
\usepackage{multirow}

\def\balpha{\boldsymbol{\alpha}}
\def\b{\textbf{b}}
\def\bK{\mathbf{K}}
\def\bC{\mathbf{C}}
\def\A{\mathbf{A}}
\def\bpsi{\boldsymbol{\psi}}

\definecolor{darkgreen}{rgb}{0,0.5,0}

\newcommand{\rev}[1]{{#1}}

\begin{document}

   \title{Optimization of starshades: focal plane versus pupil plane}

\author{R. Flamary\inst{1}
\and C. Aime   \inst{1}}

   \institute{ Laboratoire Lagrange, UMR CNRS 7293, Universit\'e de Nice Sophia-Antipolis, Observatoire de la C\^ote d'Azur,  Parc Valrose, 06108 Nice, France \\
\email{remi.flamary@unice.fr, claude.aime@unice.fr}\\}

   \date{\today}

  \abstract
       {}
     {We search for the best possible transmission for an
     external occulter coronagraph that is dedicated to the direct observation
     of terrestrial exoplanets. We  show that 
     better observation conditions are obtained when the flux in the
     focal plane is minimized in the zone in which the exoplanet is
     observed, instead of the total flux received by the telescope. }
      {We describe the transmission of the   occulter as a sum of
      basis functions. For each element of the basis, we numerically
      computed  the Fresnel
      diffraction at the aperture of the telescope and the complex
      amplitude at its focus. The  basis
      functions are circular disks that are linearly apodized over a few
      centimeters (truncated cones). We complemented the numerical
      calculation of the
      Fresnel diffraction for these functions by a
      comparison with pure circular discs (cylinder) for which an
      analytical expression, based on a decomposition in  Lommel
      series,  is available. The technique of deriving the optimal
      transmission for a given spectral bandwidth is a
      classical regularized quadratic minimization of intensities, but
      linear optimizations can be used as well.
}
       {Minimizing the integrated intensity on the aperture of the
     telescope or for selected regions of the focal plane leads to
     slightly different transmissions for the occulter. For the focal
     plane optimization, the resulting residual intensity is
     concentrated behind the geometrical image of the occulter, in a
     blind region for the observation of an exoplanet, and the level
     of background residual starlight becomes very low outside this
     image.\rev{ Finally, we provide a tolerance analysis for the
       alignment of the occulter to
     the telescope which also favors the focal plane optimization.This
     means that telescope offsets of a few decimeters do not strongly reduce the efficiency of the occulter.}
}
     {}

\titlerunning{Optimization of starshades}

   \maketitle

\section{Introduction}
  
Almost twenty years ago  \cite{1995Natur.378..355M} made the first
detection of an exoplanet, and more than a thousand  are now
recognized\footnote{see for example \url{http://exoplanet.eu/}}. External occulters
will be formidable instruments for performing detailed spectral \rev{
  analysis}  of exo-Earths that have  previously been detected by other methods. 
   
   The first use of an external occulter  dates back to
   \cite{1948JOSA...38.1083E}, who used an occulting disk set in front
   of a  very small   coronagraph of  \cite{1939MNRAS..99..580L}.
   Very soon solar astronomers  have identified  diffraction effects
   of a circular occulter and considered various kinds  of apodized,
   shaped, or multiple occulters for space missions, as  well described
   in the review paper  of  \cite{1988SSRv...47...95K}. 
 Because of implementation difficulties,  toothed  or multiple discs are
 preferred over radially graded ones, which have been described in
 \cite{purcell1962coronagraph} and \cite{newkirk1965coronascope}. Next
 to the successful Solar and Heliospheric Observatory mission
 (\cite{1995SoPh..162..357B}),  future solar coronagraphy envisages
 formation-flying spacecrafts with an occulter of diameter 1.5 m at
 about 150 m in front of a small  telescope
 (\cite{vives2009aspiics}).
   
   The parameters  are completely different for  exoplanets,. In
   rounded numbers, the occulting angle is about 0.2 arcsec against
   2000 arcsec for the Sun, and a large 4m diameter telescope is
   needed with an occulter of 50 m diameter at a distance of 80 000
   km.  The positive aspect
    of exoplanet experiments is that  diffraction phenomena are less
    difficult to simulate  numerically than in the solar case. This is
    due to two reasons. There are fewer than 20  Fresnel zones compared
    with 8000 in the solar case, and the main source of
    diffracted light is  a point source instead of a huge $1/2^\circ$
    extended object. Nevertheless, the problem remains difficult
    because of the high expected dynamic \rev{ range}.

Progression of ideas for exoplanets first followed developments for
the Sun, with little interaction, but now the theoretical calculations
and instrumental developments have resulted in much more  developments
than in the solar case. In his review of space astronomy
   \cite{spitzer1962beginnings} 
   mentioned the technique of apodization envisaged by solar
   astronomers. Elaborated petal-shaped occulters can already be found
   in  \cite{1985AcAau..12..195M}, and \cite{2000ApJ...532..581C}
   proposed   a circular  apodizing transmission on a square
   mask. Studies of shaped and  apodized occulters  have been reported
   in many publications, such as \cite{Cash2006}, \cite{Arenberg2007},
   \cite{Vanderbei2007}, \cite{2010SPIE.7731E..77S},
   \cite{2011ApJ...738...76C} and \cite{2011JOSAA..28.1668W}, to
   cite just a few.    
Although partially transparent petal shaped occulters are envisaged by \cite{2013JOpt...15c5705S}, the most advanced technological developments are for petaled star shades. The  two-step procedure leading to an optimal shaped occulter is well described in  \cite{2013aero.confE.201K}. 
These authors explained that they first seek for the optimal variable   transmission of an apodized occulter
and then  chose a sufficient number of petals for the shaped occulter
to  best approach the  theoretical result given by the variable
transmission. We here focus only on
the first part of this approach, that is, on the search for an ideal
transmission, leaving  the final design  of a shaped
occulter to a future work.

Our purpose is to show that the occulter can be optimized
advantageously  from the telescope
  focal plane, which will minimize the level of light that is
  diffracted by
  the star in
  a zone of interest of the focal plane for the exoplanet. The
  resulting illumination on the telescope aperture appears to be
  apodized from center to edge. The integrated flux on the telescope
  pupil is no longer  minimized.  
 The observation  is nevertheless improved because most of
 the residual flux is trapped behind the geometrical image of the
 mask, in a blind area for the observation of the exoplanet. 
 Differences in transmission with an occulter that minimizes the flux
 over the aperture are small, but the  gain in the residual background
 light at the level of the exoplanet range from 1.9 to 50 depending on
 the spectral bandwidth,
 which might  correspond to a gain in integration time of a factor 3.6 to 2500
 for certain observing conditions, in \rev{ photon counting} mode.

The paper is organized as follows: The
 fundamental relations that describe the complex amplitudes diffracted by the
 occulter in the aperture and focal planes are given in Sect. 2. The
 procedure for optimizing the transmission of \rev{a circularly symmetric apodized occulter} to
 minimize the flux on the telescope aperture or for selected regions
 of the focal plane are described in Sect. 3. Sect. 4 contains results
 on the optimization problem and
 a discussion. Additional information is given in two appendices
based on an analytic approach of the Fresnel diffraction of a pure
circular disc  and \rev{of a linearly apodized disc}.

\section{Expressions of complex amplitudes and intensities in the aperture and focal planes}

\begin{figure*}
 \centering
\includegraphics[width=\linewidth]{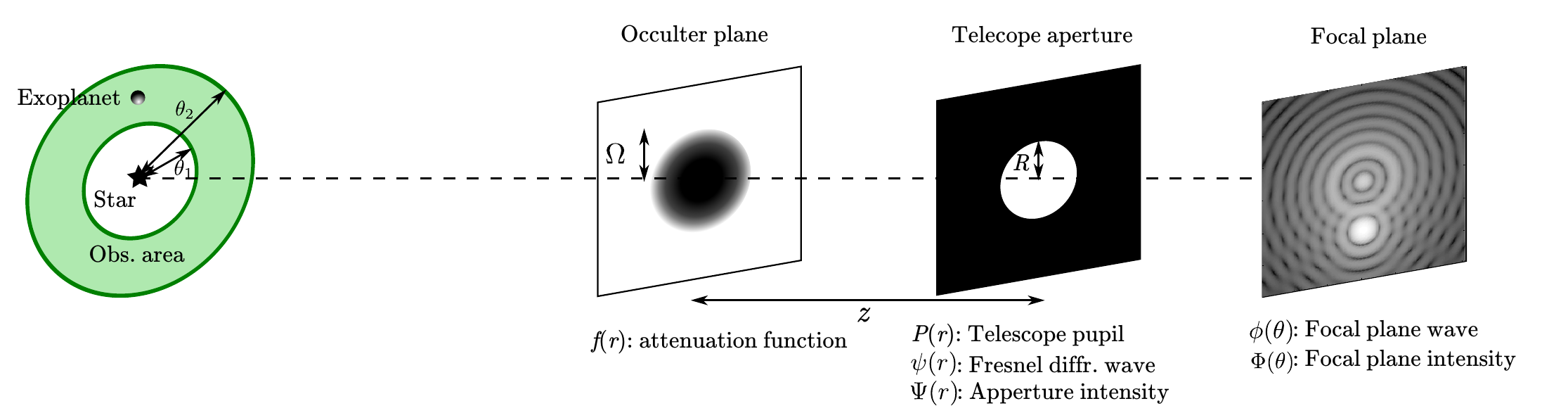} 
  \caption{ Schematic diagram of an external occulter coronagraph
    with the principal notations used in the paper. The observation,
    occulter, aperture, and focal telescope planes are shown.}
  \label{fig:schema}
\end{figure*}

We \rev{denote} with $2 \Omega$ the overall diameter of the occulter.  For  an occulting mask, it is convenient to write its  transmission as $ t(r)=1-f(r)$, \rev{with the attenuation function $f(r)$ constrained  by }  $0 \le f(r) \le 1$ for $|r| \le \Omega$,  and $f(r)=0$ for $|r|>\Omega$.   For a wave of unit amplitude,  the Fresnel diffraction at the telescope aperture at a distance $z$ from the occulter (see Fig. \ref{fig:schema}) can be written as
\begin{equation}
\label{EqBase}
\psi(r)=1-\frac{\tau_z(r) }{i \lambda z}  \int_0^{\Omega}  2 \pi  \xi f(\xi) \tau_z(\xi) J_0(2 \pi \frac{\xi r}{\lambda z}) d\xi,
 \end{equation}
 \noindent where  $\tau_z(r)= \exp(i \pi r^2 / \lambda z) $ is a quadratic phase-term corresponding to a diverging lens, and  $J_0(r)$ is the Bessel function of the first kind.

This relation is equivalent to Eq. 4  of \cite{Vanderbei2007}, neglecting here the term of propagation of a plane wave. 
  In general, this integral does not admit an analytical solution,
  except for $f(r)=1$, as shown by \cite{2013A&A...558A.138A} and
  outlined here in Appendix \ref{A1}. Therefore, Eq.\ref{EqBase}, which
  appears as the   Hankel transform of $f(r)  \tau_z(r)$, must be
  evaluated numerically.  Computations of these transforms are
  delicate, as discussed  by   \cite{lemoine1994discrete} for example,
  who recommended a sampling  of the function based on  zeroes of the
  Bessel function $J_0$. In practice, after several tests described in
  Appendix  \ref{A1} and  \ref{A2} and below, we
  directly used the function NIntegrate of \textit{Mathematica}, which
  performs very well for this kind of calculation, thanks to recent
  improvements of numerical integration.

 This wavefront $\psi(r)$ arrives onto the   aperture of the telescope
 of transmission $P(r)$, and we denote with $\varphi(r)= P(r)\;\psi(r)$ the complex amplitude of the wave going through the telescope aperture.
  For the sake of simplicity, we assume in the following a perfectly
 circular  telescope of diameter $2R$, centered on the optical axis of
 the system, so that $P(r)=\Pi(r/R)$, defining here for convenience
 that
 the box distribution $\Pi(r)$ equals  1 for $|r|<1$ and 0
 elsewhere.  Note that a more realistic telescope with a central
 obscuration can be easily inserted into the calculations. 
 Using the circular symmetry of the problem, the complex amplitude of
 the wave in the focal plane can be written as a Hankel transform of
 $\varphi(r)$ of the form:
  \begin{equation}
\label{AmpliPlanFocal}
\phi(r)=\frac{ \tau_F(r)}{i \lambda F} \int_0^{R}  2 \pi  \xi \varphi(\xi)  J_0(2 \pi \frac{\xi r}{\lambda F}) d\xi 
=\frac{\tau_F(r)}{i \lambda F}\hat{\varphi}( \frac{ r}{\lambda F}),
 \end{equation}
 \noindent where   $F$ is the focal length of the telescope. Note that
 here  the phase term $\tau_F(r)$ is compensated for by the lens phase
 function of the telescope, meaning that a simple 2D Fourier transform
 or Hankel transform denoted by $\hat{\varphi}$ substitutes the more
 complex Fresnel propagation of Eq. \ref{EqBase}.

 It is  moreover convenient to calibrate the focal plane in terms of angular units $\theta= r/F$ on the sky, and the intensity in the focal plane of the telescope can be written
 \begin{equation}
\label{IntPlanFoc}
\Phi(\theta)=\frac{1}{\lambda^2}|\hat{\varphi}( \frac{ \theta}{\lambda })|^2,
\end{equation}
  
\noindent instead of just the Airy pattern
  \begin{equation}
\label{Airy}
\Phi_0(\theta)=\frac{1}{\lambda^2}|\hat{\varphi_0}( \frac{ \theta}{\lambda })|^2=
\frac{\pi^2 R^4}{\lambda^2}\;
 \left(\frac{J_1(2 \pi \theta R/\lambda)}{\pi \theta R/\lambda} \right)^2
\end{equation}
 \noindent for the direct observation without external occulter with a perfect telescope of circular aperture of diameter $2R$. 
 
Limiting the analysis to only geometrical aspects, the occulter
prohibits the planet observation for an angular radius smaller than
$\Omega/z$ to the star, and the light coming from the planet passes
over
the occulter for an angle $\theta _0 \geqslant (\Omega+R)/z$, a
variant of the geometric inner working angle $\Omega/z$ of
\cite{2013aero.confE.201K}, taking into account the telescope
size. \rev{Note that all the wave and intensity functions defined in this section
are illustrated in Figure \ref{fig:schema}.}

 \section{Procedure used for optimizing $f(r)$ for integrated intensities}

The basic idea of  an external occulter coronagraph is to avoid
letting the stellar
light  enter the telescope while leaving the light coming
from the exoplanet unchanged .   
In a first approximation, if the  angular position of the planet is  larger than $\theta_0$, we 
can expect that the flux coming from the planet is almost unaffected by the occulter, and a fair measurement of the 
efficiency of the external occulter can be the normalized residual integrated intensity over the 
 telescope aperture, for example the quantity $\Gamma$ defined by
\begin{equation}
\label{G}
\Gamma= \frac{2 \pi}{\Delta\lambda} \int_{\lambda_m}^{\lambda_M}\int_0^R r \;  \Psi(r) dr d\lambda,
\end{equation}
\noindent where $ \Psi(r) =|\psi(r)|^2$ is the wavelength-dependent
intensity and  $\Delta\lambda$ is the spectral bandwidth of the
experiment,  simply taken equal to  $\lambda_M-\lambda_m$ here. 
In this analysis, the same weight is assigned to each wavelength,
independently of the brightness of the source, the quantum
efficiency of the detector, or the presence of acquisition filters (\cite{2010SPIE.7731E..77S}).  It would be easy to perform a differently
weighted average for a more realistic analysis, if necessary. Our
numerical computation has been made from $\lambda_m=$ 380 nm to
$\lambda_M=$ 750 nm and with $R=$ 2 m.

To find an optimal shape  for the external
  occulter, a natural approach would be to minimize $\Gamma$ with
  respect to $f$. Note that this approach has been used
  in \cite{2011JOSAA..28.1668W}, but this is not the
  only way to optimize the shape of an occulter. For instance,
  \cite{Vanderbei2007} and \cite{2013aero.confE.201K} proposed to minimize the
  upper bound $c$ on the real and imaginary part of $\psi(r),\forall
  r\in[0,R]$.  This approach will ensure that the intensity in the
  aperture plan will be below $2c^2$ and also minimize the
  flux. We \rev{emphasize } that while we decided here to
  minimize the flux, our approach can be readily adapted to the
  minimization of the upper bound instead of the whole flux $\Gamma$.

Because of the conservation of energy,  the  flux is conserved from aperture to focal planes. Nevertheless, we are in fact
  interested in the
level of background light produced by the star in the area where we
observe the planet. A quantity representative of the residual flux may
be given by the residual light $\gamma$ over a zone of the focal
plane, such as
\begin{equation}
\label{g}
\gamma=  \frac{2 \pi}{\Delta\lambda} \int_{\lambda_m}^{\lambda_M} \int_{\theta_1}^{\theta_2} \theta \; \Phi(\theta) d\theta d\lambda,
\end{equation}
\noindent  where $\theta_1$ and $\theta_2$ define the region of
interest of the observation (Fig. \ref{fig:schema}), corresponding for example to the habitable zone around the star, or the possible excursion of a known planet over a star, with probably
 $\theta_1$  close to $\theta_0$.
 We therefore have $\gamma \le \Gamma$, the equality holding in the limit $\theta_1=0$ and $\theta_2=\infty$.

The two measures of residual light $\Gamma$  and $\gamma$ are
  based on different \rev{points} of view. If one manages to completely
 \rev{ shut off } the light in the aperture, then the residual light in the
  observation area will also \rev{shut off}, but the reverse is false. In
  other words, minimizing $\Gamma$ or $\gamma$ with respect to $f$ will
lead to different optimal solutions. We believe that minimizing
$\gamma$ is better when the objective is to observe an exoplanet in a
known area of the focal plane.

\subsection{Decomposition of $f(r)$ on a basis of functions $f_k(r)$}
\label{sec:flux-optim-probl}

As already presented in \cite{jacquinot1964ii} for the systematic search for apodizing properties, a common way to optimize the shape of a function is to force this
function to be a weighted sum of basis functions. We here aim
to optimize the weights $\alpha_k$ such that
\begin{equation}
  \label{eq:occult_func}
  f(r)\quad=\quad\sum_{k=1}^K \alpha_k f_k(r) \quad \text{and} \quad
0 \leq f(r) \leq 1.
\end{equation}

 \cite{jacquinot1964ii} have shown that, if the expansion contains an
 infinite number $K$ of terms, "\textit{the absolute optimal function
   according to the criterion chosen}" is obtained regardless of the bases used. This result was obtained in the different context of
 Fraunhofer diffraction, but given the linearity of the equations, the
 result still holds for our purpose.
 In practice, however, the number $K$ of elements of the bases must be limited and   the choice of basis functions  \rev{becomes} important.

\begin{figure}
 \centering
\includegraphics[width=1\columnwidth]{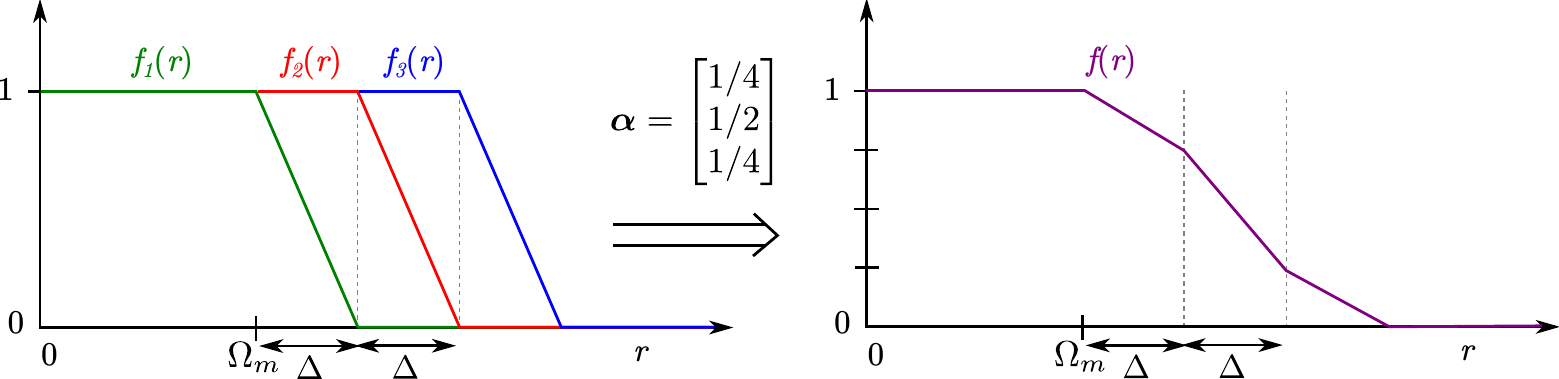} 
  \caption{ Left: illustration of the basis function used to represent
    the \rev{attenuation} function $f(r)$ with an example of the combination of three of these functions (right). For clarity of representation, the values of $\Omega_m$ and $\Delta$ are not to scale of those used for the $f_k(r)$ basis functions.}
  \label{basis}
\end{figure}

After various tests,  we used trapezoidal functions for the $f_k(r)$,  
equal to 1 for $r \le \Omega_m+(k-1) \Delta$ and equal to 0 for $r \ge
\Omega_m+k \Delta$. The functions  linearly decrease from 1 to 0
between  $\Omega_m+(k-1) \Delta$ and $\Omega_m+k \Delta$. These
function, illustrated in Fig. \ref{basis}, are very similar to the
binary disk for small $\Delta$. 
It is the same for their Fresnel diffraction pattern, as shown in Appendix \ref{A2},  but  this small difference is sufficient to lead to much better numerical results. Note
that these basis functions lead to a transmission that is piece-wise
linear as opposed to the use of binary disk that leads to piece-wise
constant transmission function.
For the numerical analysis, we set
$\Delta$ equal to 5 cm and used $K=300$ functions spreading  between
$\Omega_m= 10 $ m to $\Omega_M=\Omega=25 $ m. An example of such functions is
given in Fig. \ref{basis}. The reason for the  existence of an innermost opaque section (i.e. $t(r)=0$ or $f(r)=1$ for $r \leq \Omega_m$) is of technical origin linked with the spacecraft, as indicated for example by \cite{Vanderbei2007}.

\rev{Note that expressing function $f(r)$ as a sum of basis
  function leads to a more general model potentially at the cost of
  more complex numerical computation. For instance, one might choose to
use a more elaborated basis of functions such as prolate
functions. This would potentially lead to very good results because those
functions are known to provide efficient apodizers, and might also be excellent occulters. Nevertheless, the
physical constraints on $f(r)$ as discussed in the following would
be more difficult to express. In this particular case, it would be
advantageous to compute these constraint on a finite sampling of the
function, as in \cite{Vanderbei2007}.}

\subsection{Constraints on $f(r)$ and $f_k(r)$}
\label{sec:constr-transm-funct}

The final \rev{ attenuation} function $f(r)$ has to follow a number of physical
constraints. On the one hand, some of those constraints can be easily enforced through
a wise choice of the basis functions. For instance the smallest radius
of apodization is set by $\Omega_m$, while the largest radius is
selected by $\Omega=\Omega_m+K\Delta$. On the other hand, the \rev{attenuation}
function $f(r)$ has to be equal to $1$ for $r<\Omega_m$, and since all
$f_k(r)$ functions are equal to 1 for $r \le \Omega_m$,  we have to
enforce the
linear constraint $ \sum_{k=1}^K \alpha_k=1 $. This constraint can
also be expressed as a linear equality of the form
$\mathbf{1}^\top\balpha=1$, where $\mathbf{1}$ is a vector of ones. 

Another more
complicated constraint is the fact that $0\leq f(r)\leq 1$, $\forall
r$. These block constraints, thanks to the simple structure of the $f_k$ functions, can be expressed as a linear inequality
constraints $\A\balpha\geq \b$  with 
$$\A=
\begin{bmatrix}
  \bC\\-\bC
\end{bmatrix} \quad\bC=
\begin{bmatrix}
  1 & 1 & \dots&1 &1\\
0& 1 &\dots&1&1\\
0&\vdots &\ddots&\vdots&1\\
0& 0&\dots&1&1\\
0& 0&\dots&0&1\\
\end{bmatrix}
, \quad\text{and}\quad\b=
\begin{bmatrix}
  0\\\vdots\\0\\-1\\\vdots\\-1
\end{bmatrix}.
\quad
$$
Another constraint that has been used in
previous works is the smoothness of the transmission function. This is
typically enforced using \rev{derivatives} of the function, but in our case
smoothness can be readily enforced using a classical quadratic
regularization term $\balpha^\top\balpha$. Indeed,  minimizing the
euclidean norm of the $\balpha$ vector under the constraint $\mathbf{1}^\top\balpha=1 $ tends to promote similar values on all $\alpha_k$,
leading to a smooth transmission function (linear apodization for an infinite
regularization). This kind of regularization is equivalent to the
constraint on the curvature of $f$ described in \cite{2013aero.confE.201K}.

An additional constraint that has been enforced in previous
literature is to force $f$ to be monotonic, which in this case is a
decreasing function \emph{w.r.t.} the radius $r$. This constraint, also
enforced in \cite{2013aero.confE.201K}, has an extremely simple form in
our formulation, in effect it is the positivity constraint
$\balpha\geq \boldsymbol{0}$. To have a grasp of the
performance loss due to this additional constraint, we performed an
optimization with and without this constraint in the numerical
experiments.

Note that in this work, we made the design choice of using linear
apodization functions as basis functions for the transmission. These
choices lead to the above-mentioned physical constraints on the
weights $\balpha$. An equivalent formulation has been proposed in
\cite{Vanderbei2007} and \cite{2013aero.confE.201K}, where the basis function are
\rev{trapezoidal} functions overlapping such that the final transmission is
also piece-wise linear. \cite{2000ApJ...532..581C},
\cite{2011ApJ...738...76C} and \cite{2011JOSAA..28.1668W}, on the other
hand, used offset high-order polynomial or hyper-Gaussian functions.

Below, we present the optimization problems derived from
minimizing the intensity  $\Gamma$  in the telescope aperture and minimizing the
intensity $\gamma$  in a region of the focal plane. 
The optimal functions are denoted as $f_\Gamma(r)$  and $f_\gamma(r)$  for the optimal \rev{attenuation} \emph{w.r.t.} the
aperture plane and
 \emph{w.r.t.} a selected region of the focal plane, respectively.

 \subsection{Minimizing the integrated starlight  in the telescope
   aperture plane : $f_\Gamma(r)$ }
\label{minG}
Substituting the decomposition of $f(r)$ of Eq.\ref{eq:occult_func} into Eq.\ref{EqBase}, we obtain
\begin{equation}
\label{Decomposition}
\begin{split}
\psi(r)&=\sum_{k=1}^K \alpha_k \left(1-\tau_z(r)\int_0^\Omega
    f_k(r) \tau_z(\xi) J_0(2 \pi \frac{\xi r}{\lambda z}) d\xi\right)\\
&= \sum_{k=1}^K \alpha_k \psi_k(r),
    \end{split}
    \end{equation}
and the corresponding intensity becomes
\begin{equation}
\begin{split}
  \Psi(r)&=\left|\psi(r)\right|^2=\left|\sum_{k=1}^K \alpha_k \psi_k(r)\right|^2=\sum_{k,l=1}^{K,K}\alpha_k\alpha_l
\psi_k(r)\psi_l(r)^\dagger\\
&=\balpha^\top\bK_\psi(r)\balpha,
\end{split}
\end{equation}
with $\bK_\psi(r)=\bpsi\bpsi^\dagger$ with
$\bpsi^\top=[\psi_1(r),\dots,\psi_K(r)]$ a rank one
matrix of general coefficient
$K_{l,k}(r)=\psi_k(r)\psi^*_l(r)$, where
$\psi_k(r)$ is the Fresnel diffraction of a wave of unit
amplitude for the basis function occulter $f_k(r)$.
The flux $\Gamma$ can then be expressed as 
\begin{equation}
  \Gamma= \frac{2 \pi}{\Delta\lambda}
  \int_{\lambda_m}^{\lambda_M}\int_0^R r \;  \balpha^\top\bK_\psi(r)\balpha
  dr d\lambda=\balpha^\top\bK_\Gamma\balpha,
\end{equation}
where $\bK_\Gamma=\frac{2 \pi}{\Delta\lambda}
  \int_{\lambda_m}^{\lambda_M}\int_0^R r \;  \bK_\psi(r)
  dr d\lambda$ is \rev{a} $K\times K$ matrix integrating all the surface of
  the aperture and all the wavelength in $(\lambda_m,\lambda_M)$. The
   optimization problem is finally defined as
  \begin{align}\label{eq:optprob}
  \quad  \min_{\balpha}&\quad \quad\balpha^\top(\bK_\Gamma+\mu \mathbf{I})\balpha\\
&\quad \text{s.t.}\quad \mathbf{1}^\top\balpha=1\quad \text{and} \quad\A\balpha\geq \b,
\nonumber
  \end{align}
where $\mathbf{I}$ is the identity matrix and $\mu$ is a positive
regularization parameter. As discussed in \cite{2011JOSAA..28.1668W},
regularization is important in this case because the matrix
$\bK_\Gamma$ is poorly conditioned and would lead to difficulties when
solving the problem. Moreover, as discussed in the previous section,
the $\mu$ parameter weight the 
quadratic regularization on the $\balpha$ vector, promoting smoothness
in the transmission function. This smoothness has also been shown to be
important in practice in \cite{2013aero.confE.201K}, where an
equivalent smoothness constraint was inserted in the optimization
problem. \rev{Also note that in practice one can easily express the
  smoothness constraint of \cite{2013aero.confE.201K} in our
  optimization problem instead of using a regularization term. Since
  with the trapezoidal basis, the second-order derivatives are exactly
  computed using the finite difference between adjacent $\balpha$
  components, the corresponding constraints can be obtained by adding
  lines to the $\A$ matrix that contain finite difference operators, and
  lines to $\b$ that contain the constraint parameter $\sigma$ of  \cite{2013aero.confE.201K}.}

The optimization problem defined Eq. (\ref{eq:optprob}) is a 
constrained quadratic optimization  \cite{boyd2004convex}. The linear
constraints forbid the
use of \rev{the closed-form solution} as proposed in
\cite{2011JOSAA..28.1668W}, but there
\rev{exist} several efficient optimization strategies in the
literature, 
among which the active set approach \rev{presented in \cite{vanderbei1999interior}}.

 \subsection{Minimizing the starlight  in a region of
   the telescope focal plane: $f_\gamma(r)$}
\label{ming}
To express, the intensity in the focal plane, we used the same decomposition of the function $f(r)$ as a sum of functions $f_k(r)$ as above, the complex amplitude in the focal plane can be written as a sum of elementary function  $\phi_k(r)$, Fourier transforms of $P(r) \psi_k(r)$ according to Eqs \ref{eq:occult_func} and \ref{AmpliPlanFocal}.
 The intensity in the focal plane of  Eq. \ref{IntPlanFoc} becomes 
  \begin{equation}
\label{IntPlanFocK}
\Phi(\theta)=\frac{1}{\lambda^2}\left| \sum_{k=1}^K \alpha_k \phi_k(
  \frac{ \theta}{\lambda })\right|^2=
\balpha^\top\bK_\phi(\theta)\balpha,
\end{equation}
and the intergrated density 
\begin{equation}
  \label{eq:1}
  \gamma=  \frac{2 \pi}{\Delta\lambda} \int_{\lambda_m}^{\lambda_M} \int_{\theta_1}^{\theta_2} \theta \; \balpha^\top\bK_\phi(\theta)\balpha d\theta d\lambda=\balpha^\top\bK_\gamma\balpha.
\end{equation}
The resulting optimization problem is exactly the same that of
Eq. \ref{eq:optprob}, but with a different metric $\bK_\gamma$. 

The optimal functions $f_\Gamma(r)$ and $f_\gamma(r)$ are different because
the  solutions $\balpha$ are different.Because of conservation of energy between aperture and focal planes, these functions become identical  in the special case
$\theta_1=0,\theta_2=\infty$,  where $\bK_\Gamma=\bK_\gamma$.
Our optimization in the focal plane was performed for $\theta_1=0.1$ arcsec and $\theta_2=0.5$ arcsec.

\section{Results and discussion}
\label{sec:numer-results-disc}

\subsection{Numerical computations}
\label{sec:numer-exper}

In the following, we denote $\Psi_\Gamma(r)$ and $\Phi_\Gamma(\theta)$
the intensities obtained in the aperture and focal plane of the
telescope using an external occulter of transmission  $1-f_\Gamma(r)$
minimizing the integrated intensity $\Gamma$ and described in
Sect. \ref{minG}.  Similarly, we denote $\Psi_\gamma(r)$ and
$\Phi_\gamma(\theta)$ the results obtained using an occulter
transmission  $1-f_\gamma(r)$, as described in Sect. \ref{ming}. 

Note that while obtaining the matrices $\bK_\Gamma$ and
$\bK_\gamma$ is a computationally intensive problem, we can use heavily
parallel computing to estimate as a first step the $\psi_k(r)$ and
$\phi_k(\theta)$ functions through numerical integration. This was
performed using \emph{Mathematica}, with $K$
computational jobs  computed by  $\approx 100$ parallel
processes on eight core Intel processors. In the numerical experiments,
both $\psi_k(r)$
and $\phi_k(\theta)$ were regularly sampled on $3001$ samples from
respectively $r\in (0 ,R+1)$  and $\theta\in (0 ,0.644)$
arcsec. \rev{Note that $\psi_k(r)$ was computed outside of the
  telescope so that we can used it for the tolerance analysis in
  section \ref{sec:tol}. Moreover, the
  behavior of the different optimization strategies outside of the
  telescope aperture  is also interesting.}
The
wavelength $\lambda$ has also been sampled regularly with $21$ samples
in the visible light interval $(380,750)$ nm.
After computing
$\psi_k(r)$
and $\phi_k(\theta)$, a numeric 2D integration can be
performed to obtain the matrices $\bK_\Gamma$ and
$\bK_\gamma$. The regularization parameter $\mu$ in the optimization
problem was chosen adaptively with $\mu=\mu_0\max_{k,l}
|K_{k,l}|$ to be less dependent on the metric matrix $\bK$. $\mu_0$ was chosen by hand to allow good attenuation performance with a
limited dynamic in the solution function $f$. For the monochromatic
study, we set $\mu_0=10^{-8}$ and for the more complex study on a
wide bandwidth, less regularization was necessary with
$\mu_0=10^{-10}$ because of a more
complex problem and a better conditioning of the $\bK$ matrix.

Finally, the code for all the numerical experiments will
be freely available on the authors website to promote
reproducible research \footnote{\url{http://remi.flamary.com/soft.html}}.

\subsection{Optimization for monochromatic light}
\label{sec:optim-monochr-light}

\begin{figure}[!t]
 \centering
\includegraphics[width=\columnwidth]{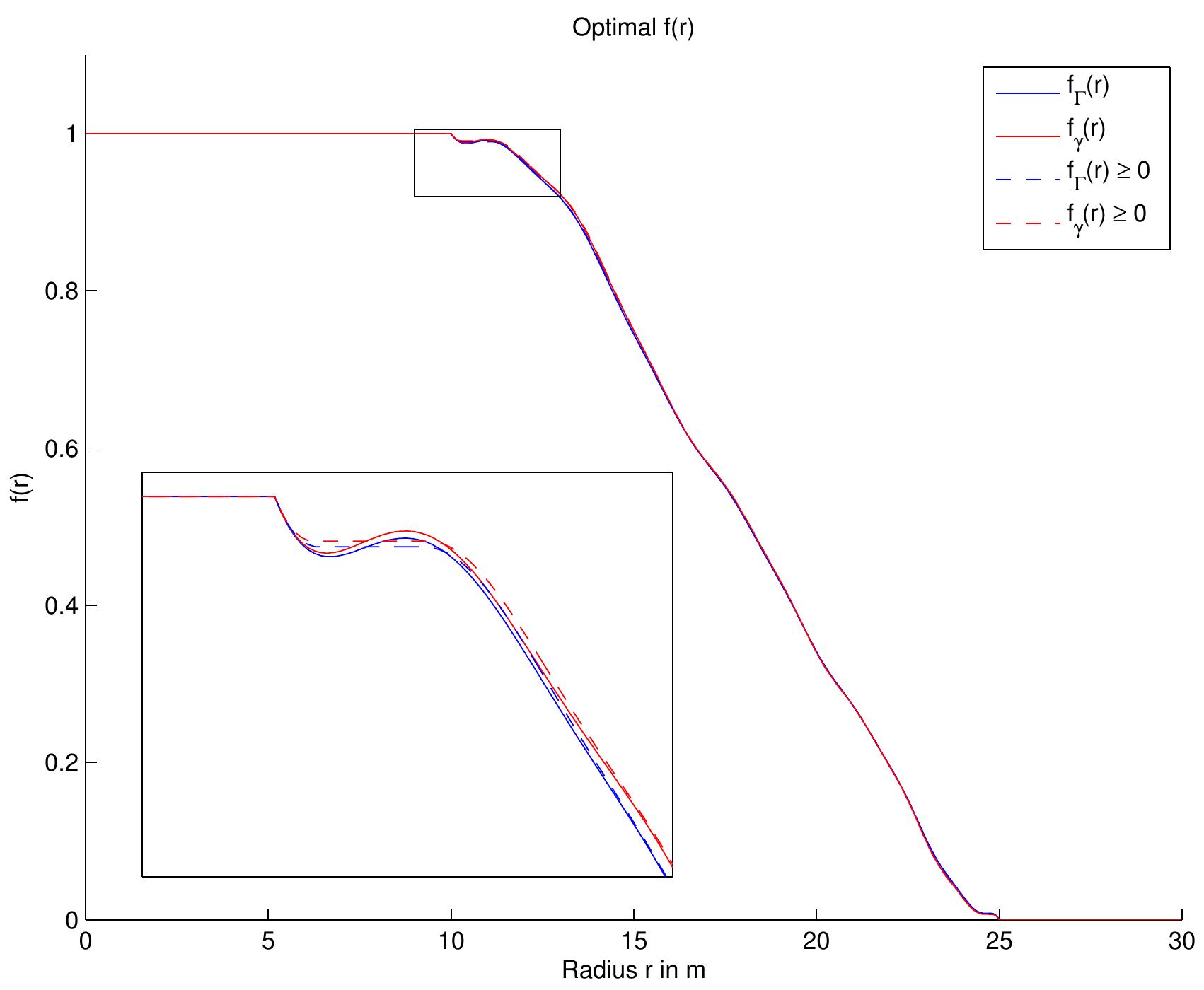} 
  \caption{ Radial cut of the \rev{attenuation} of the functions $f(r)$ for
    aperture and focal optimization at a wavelength of 562
    nm. The dashed curves denoted as $\geq 0$ correspond to an
    optimization with $\balpha\geq 0$ constraint.}
  \label{fonc}
\end{figure}

\begin{figure}
 \centering
\includegraphics[width=1\columnwidth]{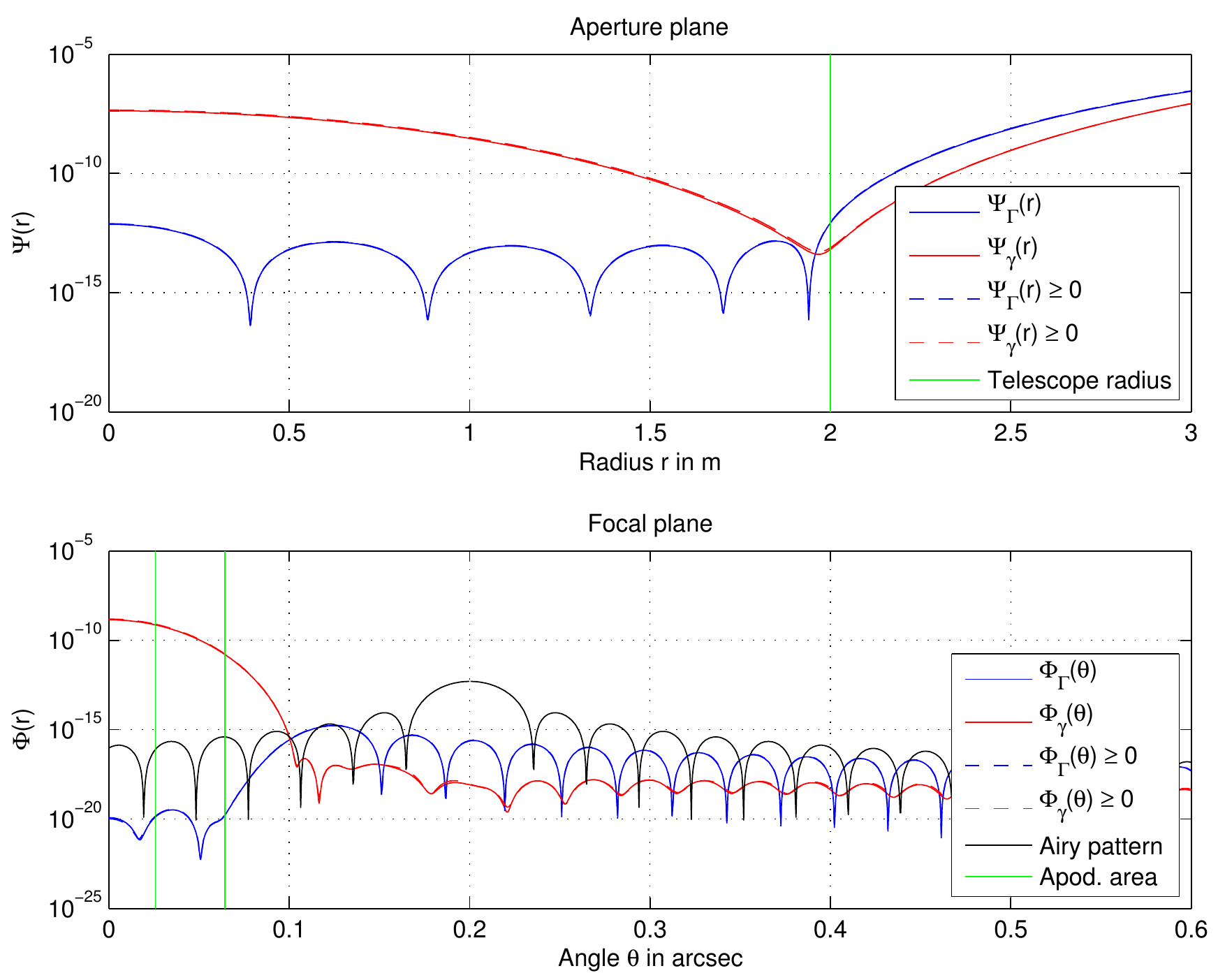} 
  \caption{  Top curves: intensity in the telescope aperture plane for
    an optimization that minimizes the integrated intensity $\Gamma$ in the aperture plane (blue curve) or $\gamma$ for optimizing the observation between 0.1 and 0.5 arcsec in the focal plane (red curve), for $\lambda$ = 562 nm. Note the apodization profile for the $\gamma$ curve \rev{and the inversion of relative position of the curves outside the region of optimization}.
   Bottom curves: focal plane intensity obtained for the two
   optimizations.  An airy pattern corresponding to a $10^{-12}$
   fainter exoplanet in
   the observation area is reported in black. The apodization area
   defined by the smallest and largest radius of the occulter is also
   reported in green. Note the inversion of behavior of curves with
   regard to the overall occulter profile.}
  \label{fonctionApod}
\end{figure}

For a single monochromatic wave, here $\lambda =$ 562 nm, we have
represented the apodizing curves $f_\Gamma(r)$
and $f_\gamma(r)$  in Fig. \ref{fonc} and the resulting  intensities  $\Psi_\Gamma(r)$ and  $\Psi_\gamma(r)$ (top curves),
and $\Phi_\Gamma(\theta)$ and $\Phi_\gamma(\theta)$ (bottom curves) in
Fig.\ref{fonctionApod}. 
We emphasize that these curves were obtained assuming a perfect
circular telescope of radius $R$. 
The $f$-curves are only slightly different and are really similar
to a linear apodization. Moreover, in the monochromatic
case, enforcing a monotonic $f$ leads \rev{to} the same results for the thickness
of the trace, which will not be the case anymore for a wide spectral
bandwidth, as we show below.

The results for the
$\Psi$ and $\Phi$-curves are quite different. 
 The residual
  starlight $\Psi_\Gamma(r)$ is  very well attenuated  across the
  telescope
  pupil, as found by  \cite{Vanderbei2007}.  In contrast,
  $\Psi_\gamma(r)$ presents a center-to-limb decrease which suggests
  an apodized structure.  
   \rev{ It is interesting to note the behavior of the curves beyond
     the region of optimization. There, the relative position of the
     curves is reversed very rapidly, the focal plane optimization
     becoming lower than the pupil plane optimization. This is an
     aspect that deserves further studies. }

  The intensity in the focal plane is also very
  interesting. While $f_\Gamma$ tends to minimize the
  intensity at all angles $\theta$, $f_\gamma$ only focuses on the region
  of interest, and a strong residual light is observed in the blind area
  behind the occulter.

The resulting
flux $\Gamma,\gamma$ for all the different optimizations is given in the upper
part of Table \ref{tab:res1lambda}.
Interestingly, $f_\Gamma$ leads to a flux $\Gamma$ in the aperture
plane $10^5$ times fainter than $f_\gamma$. Nevertheless, in the focal
plane the flux $\gamma$ in the observation area is $\approx 50$ times fainter
when using $f_\gamma$. This clearly shows the importance of using focal
plane optimization for optimal apodization.

\subsection{Optimization for a wide spectral bandwidth}
\label{sec:optim-monochr-light}

\rev{Next we investigated the two optimization strategies on a wide
bandwidth. For this we compute a regular sampling in the interval $\lambda\in$ (380,750) nm and computing the
average flux across $\lambda$. Note that both a sampling of $20$ and
$100$ values were investigated, and while we used in the
experiments a sampling of $100$ for a better precision, the final
results were very similar. In practice the variation of the
wavefront is extremely smooth \emph{w.r.t.} $\lambda$ and the selected
sampling in sufficient. }

\begin{figure}
 \centering
\includegraphics[width=1\columnwidth]{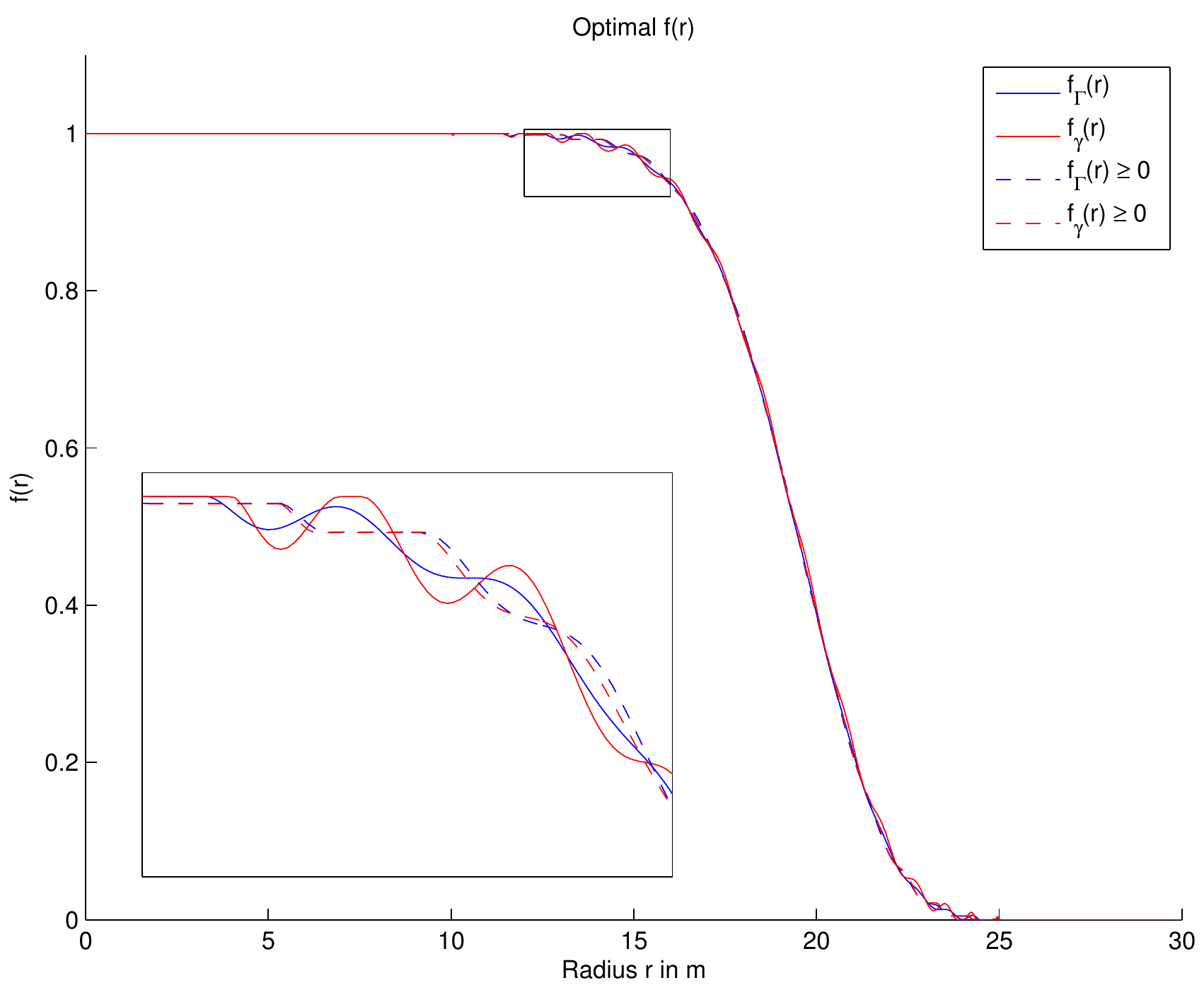} 
  \caption{ Radial cut of the \rev{attenuation} of the functions $f(r)$ for aperture and focal optimization  for the whole spectral bandwidth between 380 nm and 750 nm.}
  \label{foncb}
\end{figure}

For multiple wavelengths, $\lambda\in$ (380,750) nm, the results are
very different. First the optimal functions are much more similar,
as illustrated in Figure \ref{foncb}. The functions also tend to
oscillate and the monotonic variants are this time different from the
free variant of the optimal functions.

\begin{figure*}
 \centering
\includegraphics[width=.8\linewidth]{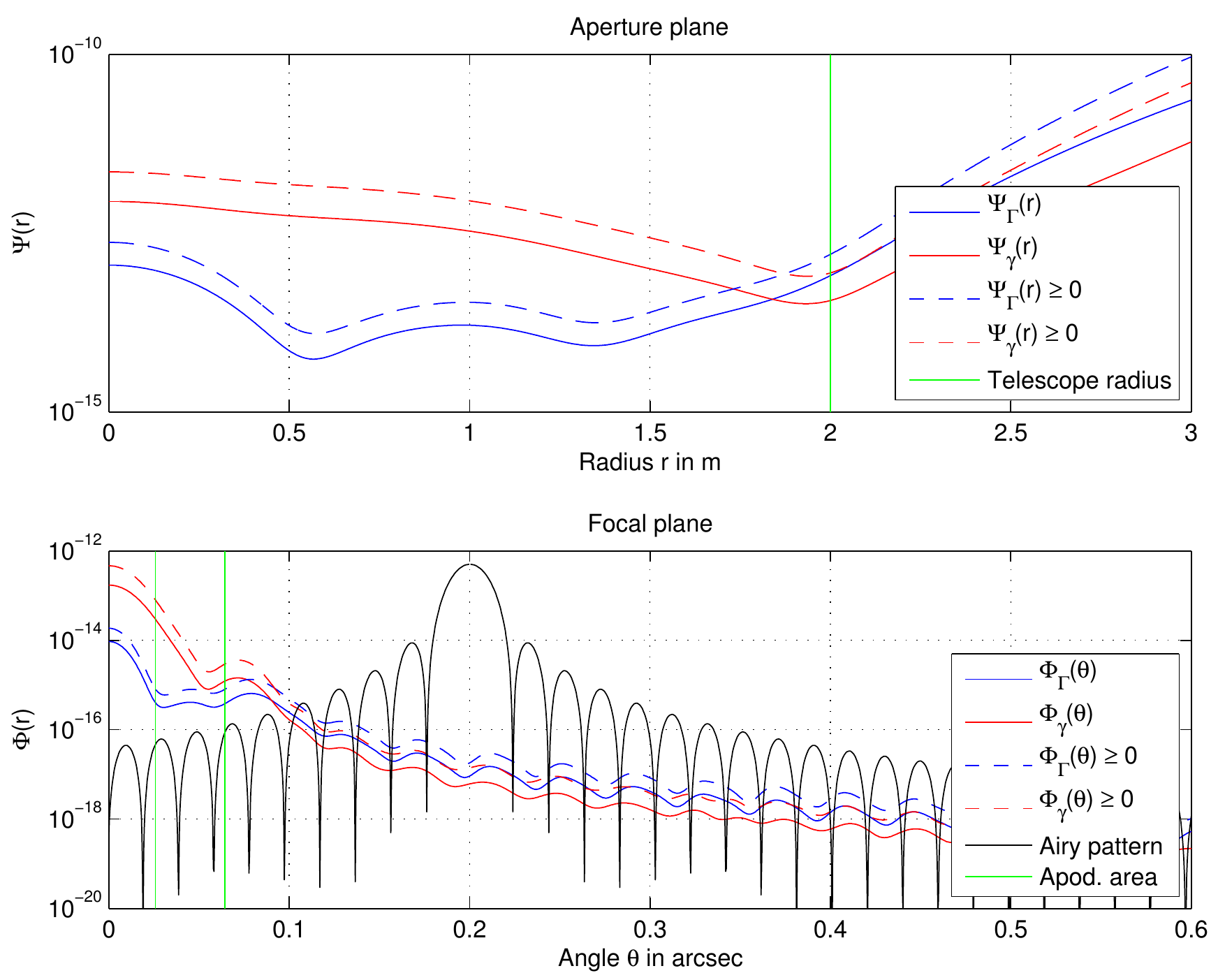} 
  \caption{ Same representation as in Fig. \ref{fonctionApod}, but for an optimization made for the whole spectral bandwidth between 380 nm and 750 nm.
  The  behavior of the inversion of the curves is  similar to the monochromatic case, although it is somewhat reduced by the effect of the chromatism.}
  \label{fonctionApodb}
\end{figure*}

The resulting intensities in the aperture and focal plan are plotted in Figure
\ref{fonctionApodb}. In the aperture plane, the intensity is still
apodized for $f_\Gamma$, but the flux is much more attenuated than in
the monochromatic case. In the focal plane, the intensities retain the
same tendencies, but the gain due to the focal plane
optimization is lower. Interestingly, enforcing a
monotonic function through $\balpha \geq 0$ leads to a clear loss
in terms of attenuation for both $f_\Gamma$ and $f_\gamma$. 

In \rev{ terms } of
quantitative performances, Table
\ref{tab:res1lambda} (lower part) shows that the flux in the observation
area of the focal plane is still more attenuated with  $f_\gamma$ with
a gain of $1.9$ in the chromatic case ($1.6$ with positivity constraint). While the gain in performance is smaller than in the
monochromatic case, it is far from negligible when observing faint
objects around a star. Interestingly, in the chromatic case, the 
 $\balpha\geq 0$ constraint leads to a loss of $2$ in attenuation (for
 both  $f_\Gamma$ and $f_\gamma$), proving that relaxing the
 constraint can lead to additional gain as long as the apodization is
 physically realizable.

\begin{figure}
 \centering
\includegraphics[width=\columnwidth]{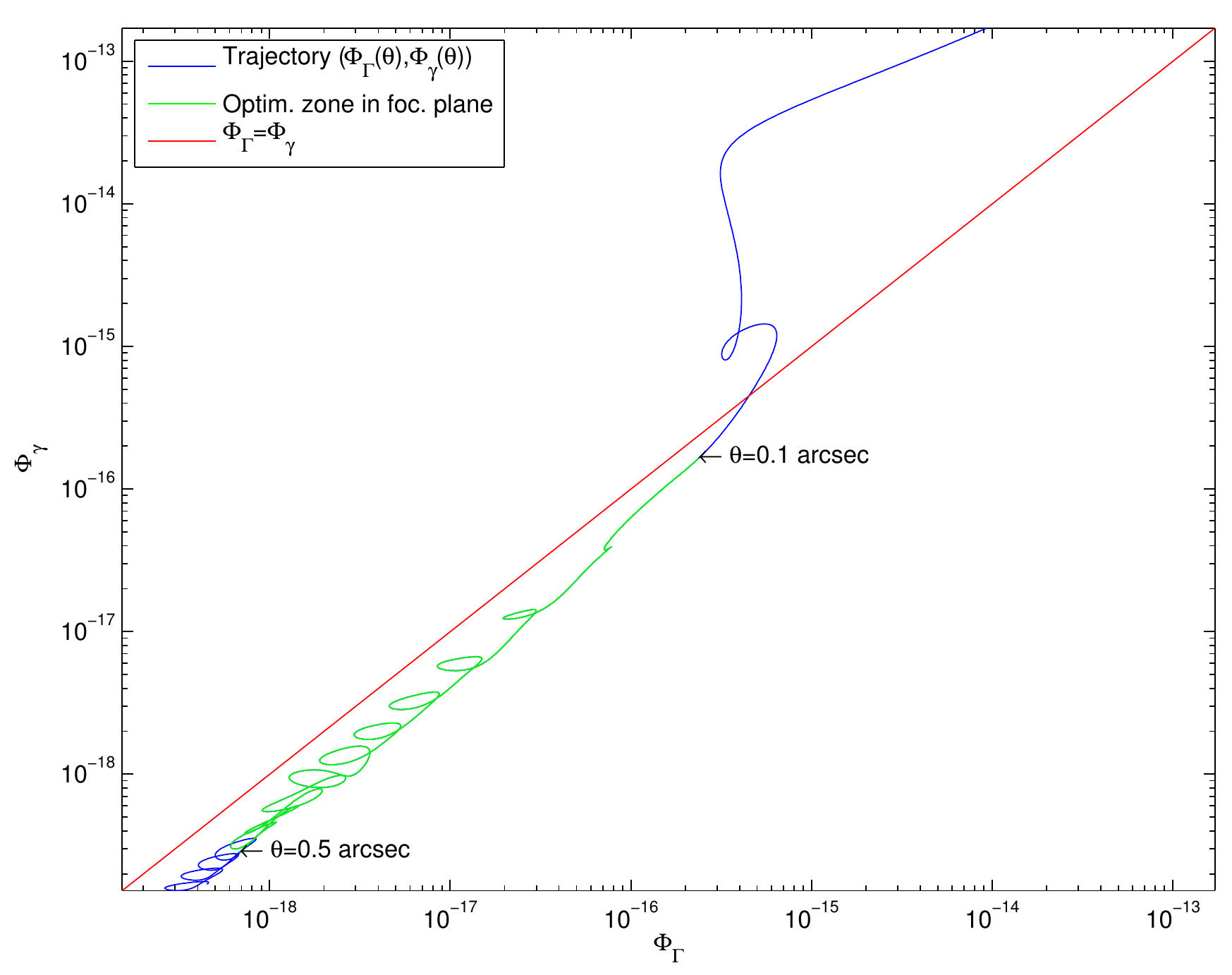} 
  \caption{ Parametric plot of the observed intensities in the focal
    plane: x-axis,   aperture plane optimization, y-axis, focal plane
    optimization.  The curve is
      drawn for the wide spectral bandwidth without positivity
      constraints. From top-right to
    bottom-left the curve follows the increase of $\theta$, clearly showing
    the concentration of light behind the occulter for the focal plane
    optimization. The line $y=x$ clearly shows that the $\gamma$-curves
   always give  better results than the  $\Gamma$-curves, especially in the
  observation area (reported in green).}
  \label{xyplot}
\end{figure}

An additional visualization comparing the
performances of $f_\Gamma$ versus $f_\gamma$ in the focal plane is
given in Fig. \ref{xyplot}. The intensity $\Phi_\gamma(\theta)$ is
plotted as a function of  $\Phi_\Gamma(\theta)$ for $\theta\in (0,  0.7)$
arcsec. The red line allows a comparison between the methods; if a
curve is below this line, it means that
$\Phi_\gamma(\theta)<\Phi_\Gamma(\theta)$ and \emph{vice versa}. We
clearly see that $\Phi_\gamma(\theta)>\Phi_\Gamma(\theta)$ for
$\theta<\theta_1$ in the blind area behind the occulter and that
$\Phi_\gamma(\theta)<\Phi_\Gamma(\theta)$ $\forall \theta \in
(\theta_1,\theta_2)$ in the observation area of the focal plane
(illustrated in green in the figure).

\begin{table*}
  \centering
  \caption{Resulting light flux in the aperture ($\Gamma$) of focal
    plane ($\gamma$) for all the optimization schemes. The upper part
    of the table corresponds to an optimization on a unique wavelength
  $\lambda$, whereas the lower part of the table corresponds to an
  optimization on a large spectrum. }
  \begin{tabular}{|c|c|cc|cc|cc|}\hline
 $\lambda$ & Flux & $f_\Gamma(r)$ & $f_\Gamma(r)$  $\geq 0$ & $f_\gamma(r)$  &
$f_\gamma(r)$ $\geq 0$ & Flux ratio & Flux ratio $\geq 0$\\\hline\hline
\multirow{2}{*}{$\lambda$=562 nm}   &$\Gamma$ & 9.07e-13 & 9.55e-13 & 5.06e-08 & 5.52e-08 & 1.79e-05 & 1.73e-05\\ 
&$\gamma$ & 7.66e-13 & 8.06e-13 & 1.38e-14 & 1.44e-14 & 5.57e+01 & 5.61e+01\\ \hline

\multirow{2}{*}{$\lambda\in$(380,750) nm}  &$\Gamma$ & 2.91e-13 & 5.88e-13 & 2.63e-12 & 7.01e-12 & 1.11e-01 & 8.38e-02\\ 
&$\gamma$ & 7.08e-14 & 1.42e-13 & 3.74e-14 & 8.99e-14 & 1.89e+00 & 1.58e+00\\ \hline
  \end{tabular}
  \label{tab:res1lambda}
\end{table*}

\subsection{Tolerance analysis to positioning errors of the telescope}
\label{sec:tol}

\begin{figure}
 \centering
 \begin{tabular}{ccc}
 \includegraphics[width=0.4\columnwidth]{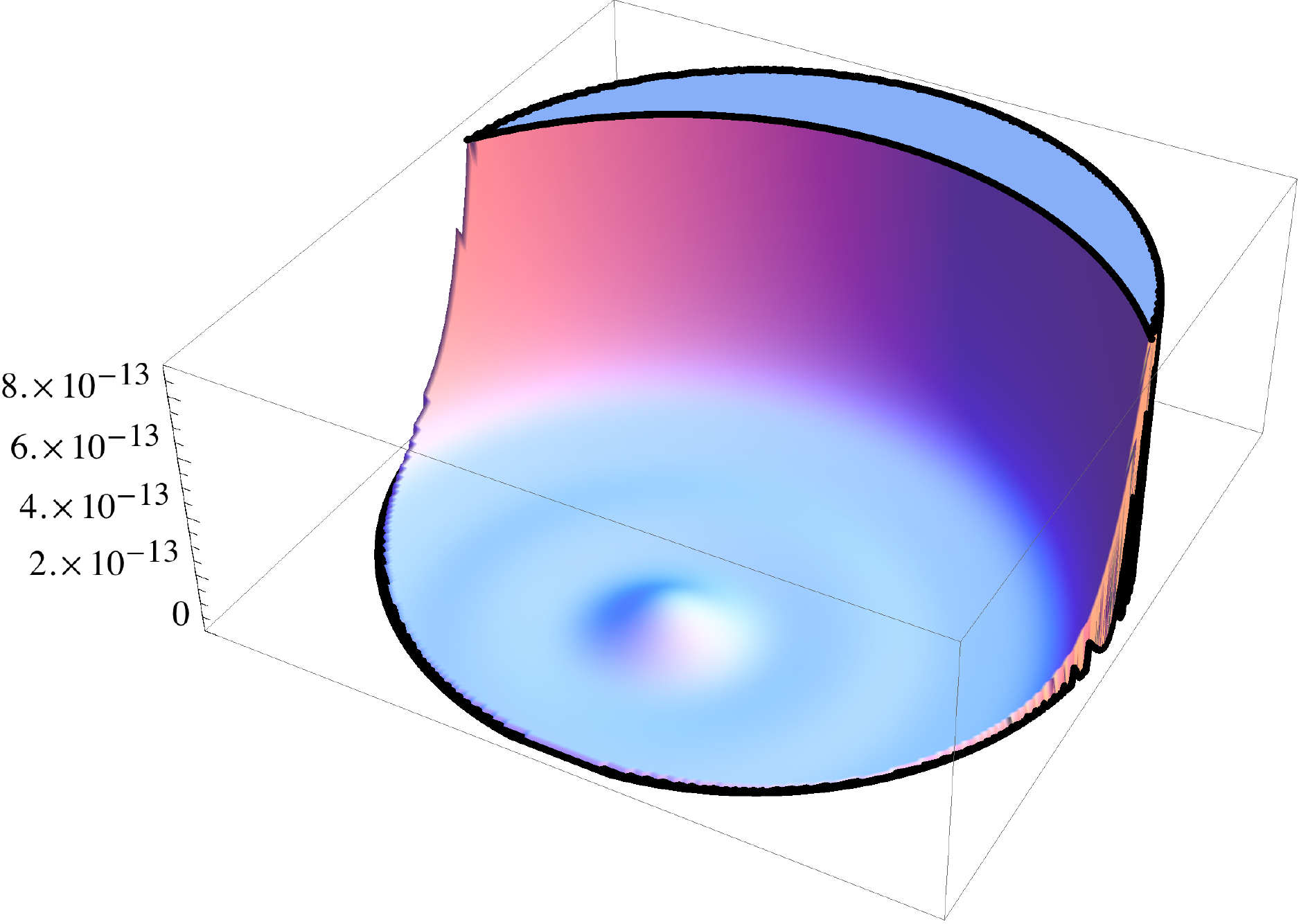}
 &
\includegraphics[width=0.4\columnwidth]{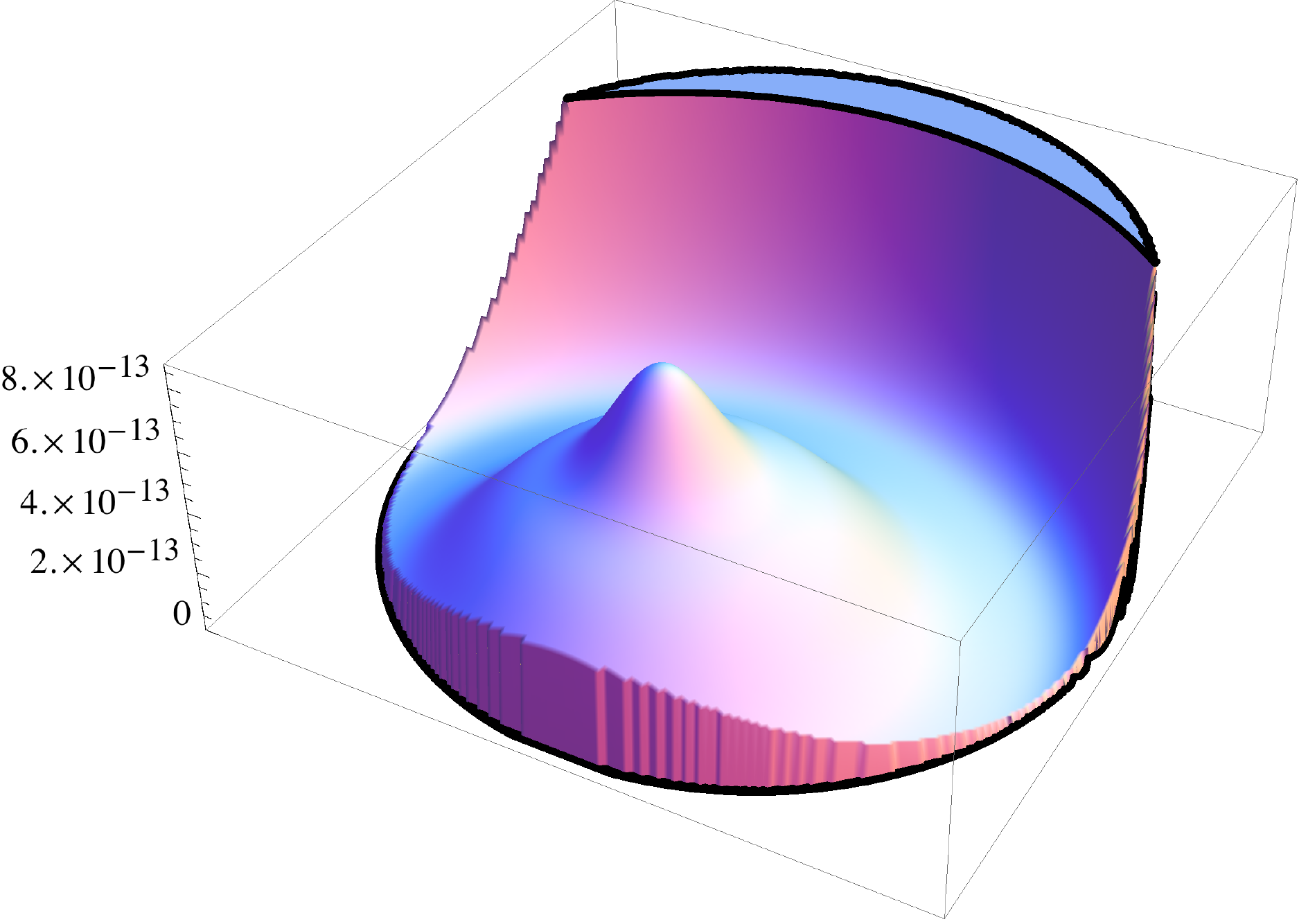} \\

\includegraphics[width=0.4\columnwidth]{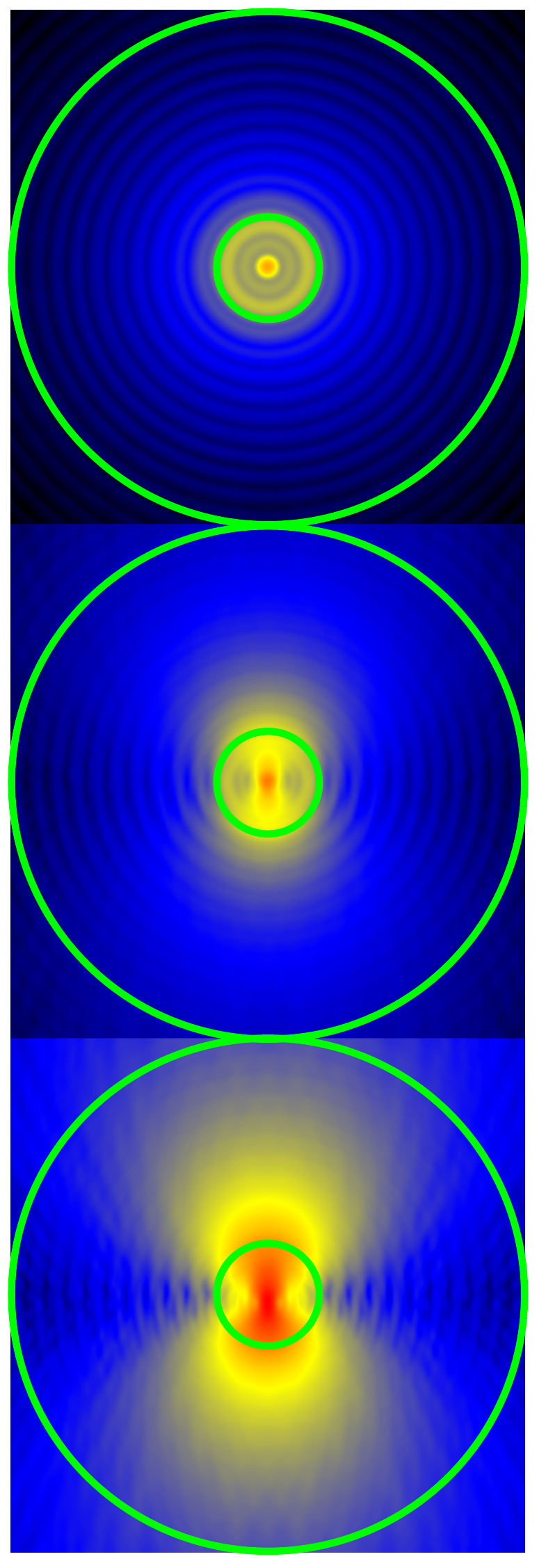} &
\includegraphics[width=0.4\columnwidth]{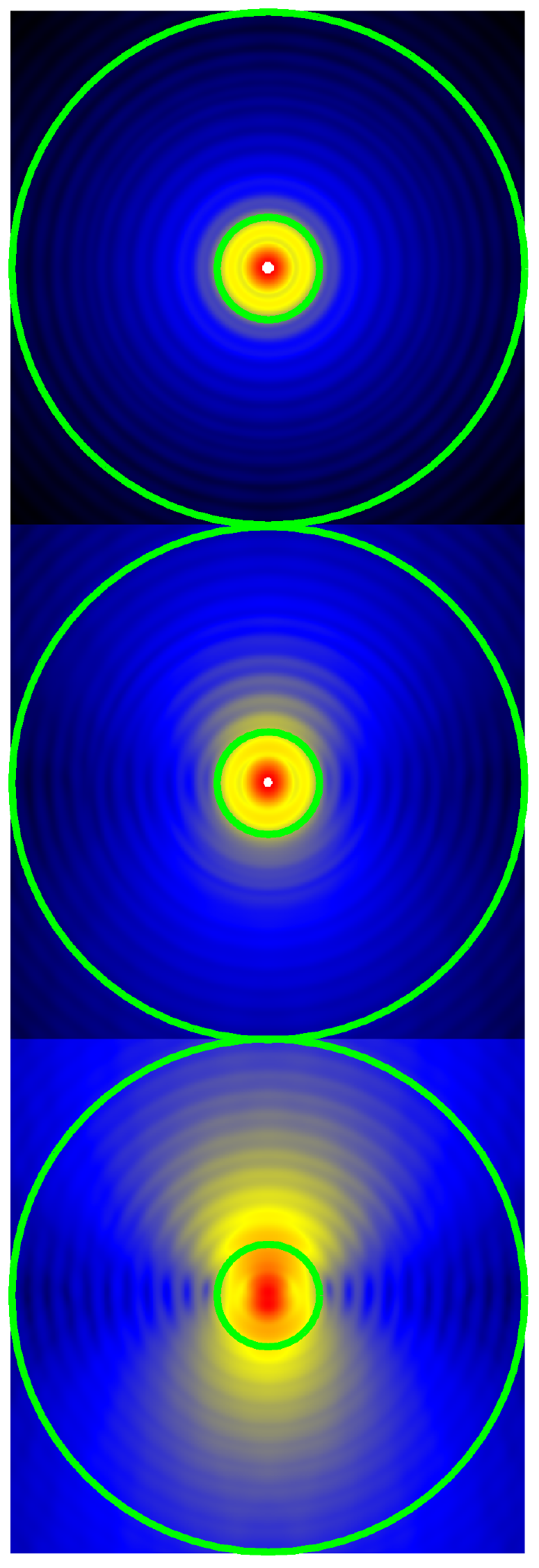} &
\includegraphics[width=0.1\columnwidth]{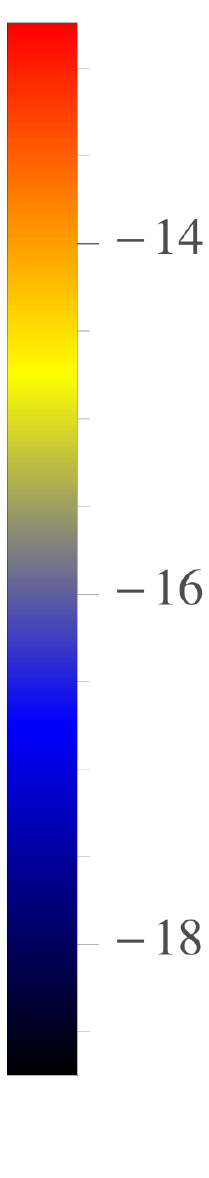}  
\end{tabular}
  \caption{ \rev{
    Top curves: pupil plane intensities for an offset of 1 m for   $\lambda\in$ (380,750) nm. Density curves below represent the focal plane intensities (log scale) for telescope positions: on-axis (upper curves) and offsets of   50 cm (intermediate curves) and  1 m (lower curves). Left: aperture optimization, right: focal plane optimization. } }
  \label{IllustrePoly}
\end{figure}

\begin{figure}
\centering
 \includegraphics[width=0.8\columnwidth]{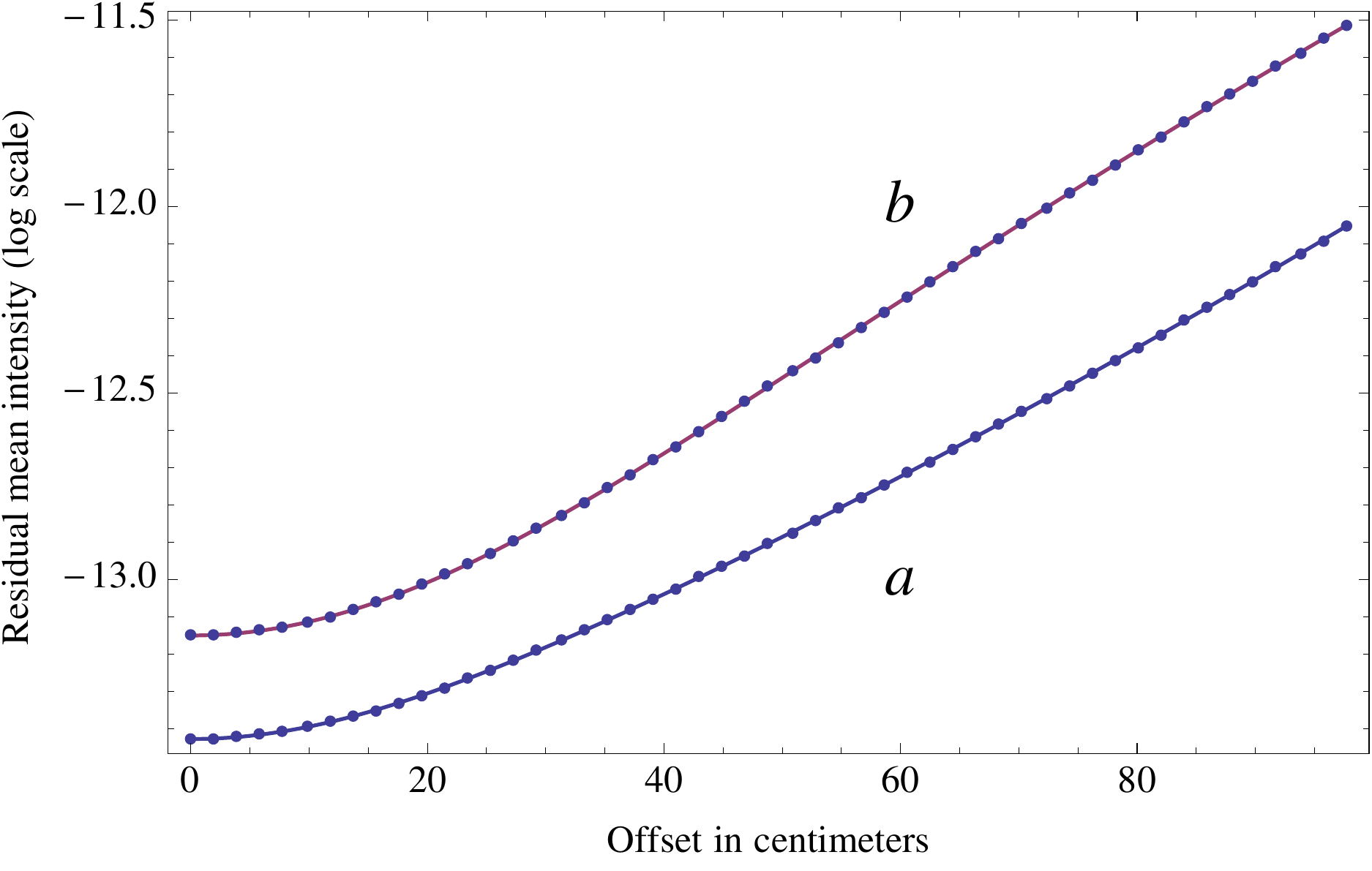} 
   \caption{\rev{Average residual light level $\gamma$ for the focal
       plane (a) and pupil plane (b) optimization, as a function of the telescope offset in centimeters  for   $\lambda\in$ (380,750) nm. The average is computed over the region from 0.1 to 0.5 arcsec delimited in  Fig. \ref{IllustrePoly} by two circles.  }
      }
  \label{PolyMoy}
\end{figure}

\rev{
A  study of the tolerance sensitivity to positioning errors for a 4m circular telescope out of its on-axis position was carried out up to an offset of 1 m. 

Because of this offset, the resulting pupil plane and focal plane
intensities are no longer circularly symmetric, and the computation of
the final focal plane image  can no longer use the Hankel
transformation of Eq.\ref{AmpliPlanFocal}. The focal plane image must
be computed taking the modulus squared of the two-dimensional Fourier
transform of the complex amplitude of the wave on the telescope
aperture.

In practice, the  study was made using a discrete Fast Fourier
Transform of an array of $1024 \times 1024$ points corresponding to
an overall  zone  of $20 \times 20$ meters,  inside which the
telescope of diameter 2m was defined by 1 or 0 transmitting pixels
over about 205 points in diameter. As a result, altogether   33 081
points were set equal to 1. This percentage of zero-padding is enough
to obtain a satisfactorily sampled focal plane image.

Positioning errors were simulated by moving the complex amplitude
obtained in the shadow of the occulter over the telescope aperture in
one direction,  by steps of 1 pixel or about 2 cm.  The
two-dimensional array was filled using  an interpolation of the
one-dimensional  complex amplitude of the wavefront computed with
Eq. \ref{EqBase}.

The focal plane image was obtained taking the modulus squared of the
Fourier transform of the wave on the telescope aperture. The effect of
wavelength was taken into account afterward. The diffraction pattern
increases in size with the wavelength. For that, a two-dimensional
interpolation of the result was used to resample the image.
Moreover,  a scaling factor in intensity inversely proportional to the
square of the wavelength was applied, as in
Eq.\ref{Airy}. As a result, the integrated flux in the focal plane
becomes wavelength independent and equal to that crossing through the
telescope aperture. As before, for the sake of simplicity, we
assumed that the received flux is constant over the spectral
bandwidth. 

Figures \ref{IllustrePoly}  and  \ref{PolyMoy}  present  results
obtained for transverse displacements of the telescope up to 1 m  from
its nominal position. We recall that the occulter shapes optimize the
measurement when the telescope is at its nominal
on-axis position, either for the integrated intensity on the telescope
aperture, or for a region in the focal plane  between 0.1 and 0.5
arcsec. In both cases the computation was made  for  the whole
bandwidth  of $\lambda\in$ (380,750) nm.

Three-dimensional representations of  the normalized intensities in
the telescope aperture were used  to enhance  differences of
intensities in the results of the two optimizations, and the
apodization-like pattern obtained for the focal plane
optimization. Both curves were clipped to make  the central
parts of the images  well visible.
As expected, there is much less light for  the aperture plane
minimization when the telescope is at its nominal position. Since the
optimization was calculated for a 4m diameter telescope, there is a
rise in the intensity at the edges for this large offset. This
increases becomes somewhat surprisingly  in favors of the  focal plane
optimization for a large off-set,  as already discussed in
Fig. \ref{fonctionApodb}.

In  Fig. \ref{IllustrePoly}   focal plane intensities are also represented in color levels, with a common color scale for the telescope on-axis and off-axis of 50 cm and 1m. 
The two circles of radius $\theta_1 =0.1$ arcsec   and $\theta_2 =0.5$ arcsec define the angular area  inside which the residual intensity is integrated.
The focal plane optimization is found to always be better than the
aperture plane optimization. From comparing these  curves, it
is  clear that the residual diffracted light remains concentrated longer  inside the first circle for the focal plane optimization, which is especially  clear  for the 50 cm offset images. This behavior is confirmed  by Fig. \ref{PolyMoy} which represents the integrated intensity in this region with a logarithmic scale. 
}

\section{Conclusion}
\label{sec:conclusion}

\rev{We addressed the problem of  optimizing of the shape of a
  circularly symmetric apodized occulter. To this end, we proposed a
  general expression for the attenuation function, as a weighted sum
  of basis functions. We expressed the optimization problem as the
  minimization of the flux, first in the pupil plane, as is commonly
  done in the literature, and then in the focal plane of a 4.m diameter
  telescope, for a region between 0.1 to 0.5 arcsec in which the
  exoplanet is supposed to be observed.

Numerical experiments show that the focal plane optimization leads to a better attenuation of the starlight than the pupil plane optimization. 
One interesting result is that while the focal plane optimization
leads to a strong flux in the pupil plane, most of this light is concentrated in the area behind the occulter in the focal plane, which is not a problem for a perfect telescope.
 Finally, the robustness of our approach to lateral misalignment of
 the occulter was investigated and showed that  the focal plane optimization also yields a better robustness.

As already indicated,  the optimization study was made for a 4m telescope assumed to be at its on-axis position. A simple procedure  to make the experience less sensitive to defect position would be to perform the optimization for a larger aperture  than is actually used. A better procedure would be  to estimate the statistics
of pointing errors and include them in the optimization procedure. Then a better tolerance to pointing errors would  be obtained, probably at the expenses of a tolerable loss in the starlight rejection. Nevertheless, our analysis showed that telescope offsets of a few decimeters will not strongly reduce the efficiency of the occulter.

This conclusion was obtained for the condition that the telescope has
a perfectly circular aperture and another study is required to model a
telescope with central obscuration. Moreover, it would be interesting to study the effect of a a more realistic telescope with small defaults of phase and amplitude. 
Deviations from this perfection could lead to somewhat different results, but the exigency of quality is limited to  a maximum of 0.5 arcsec, or about twenty Airy rings for a 4m diameter telescope. Finally, with the knowledge of the exact PSF of the telescope, one can readily adapt our optimization approach. A comparison of pupil vs focal optimization for the maximum-magnitude minimization as in \cite{2013aero.confE.201K} would also be interesting.
}

\appendix

\section{Quality test of the numerical computation using an analytical expression  for a sharp-edged circular occulter }
\label{A1}
 \cite{2013A&A...558A.138A} has shown, using an approach similar to
 \cite{born1999principles}, that an analytic expression  using Lommel series could be derived for the Fresnel diffraction of a sharp circular occulter of the form  $f(r)=f_\bullet(r,\Omega)=\Pi(r/\Omega)$, defining here $\Pi(r)$ as the function equal to 1 for $|r|\leq1$ and 0 otherwise. 
 At the wavelength $\lambda$, the complex amplitude
 $\psi_\bullet(r,\Omega)$ of the wave at the distance $z$  for
 
  (i) inside  the geometrical umbra ($r<\Omega$) and 
  
  (ii) outside ($r>\Omega$), 
  
  \noindent can be written using  two series:
 \begin{equation}
\label{Lommel}
\begin{split}
\psi_\bullet(r,\Omega) =&\\
(i)\;\;&    \tau(r) \; \tau(\Omega) \; \sum_{n=0}^\infty (-i)^n (\frac{r}{\Omega})^{n} J_{n}( \frac{2 \pi \Omega r}{\lambda z})\\
(ii) \;&1- \tau(r) \;  \tau(\Omega) \; \sum_{n=1}^\infty (-i)^n (\frac{\Omega}{ r})^{n} J_{n}( \frac{2 \pi \Omega r}{\lambda z}) .
\end{split}
 \end{equation}
At  $r=\Omega$,  both  series  converge to the same  value given by
\begin{equation}
\psi_\bullet(\Omega,\Omega) =\frac{1}{2} (1+\exp(i \frac{2 \pi \Omega^2}{ \lambda z})  J_{0}( \frac{2 \pi \Omega^2}{ \lambda z}) ).
 \end{equation}
These series
are alternative series for the real and imaginary terms, and according
to the Leibniz estimate, an upper bound of the error for
the sum limited to $n=N$ terms is given by the absolute value
of the $N + 1$  term. 
The convergence is easy when  $r/\Omega$ is smaller than 1, for the first sum for $r<\Omega$ inside the geometrical umbra, or the inverse outside.   The main error occurs near the transition zone $r=\Omega$.

\begin{figure}
 \centering
 \begin{tabular}{cc}
\includegraphics[width=0.8\columnwidth]{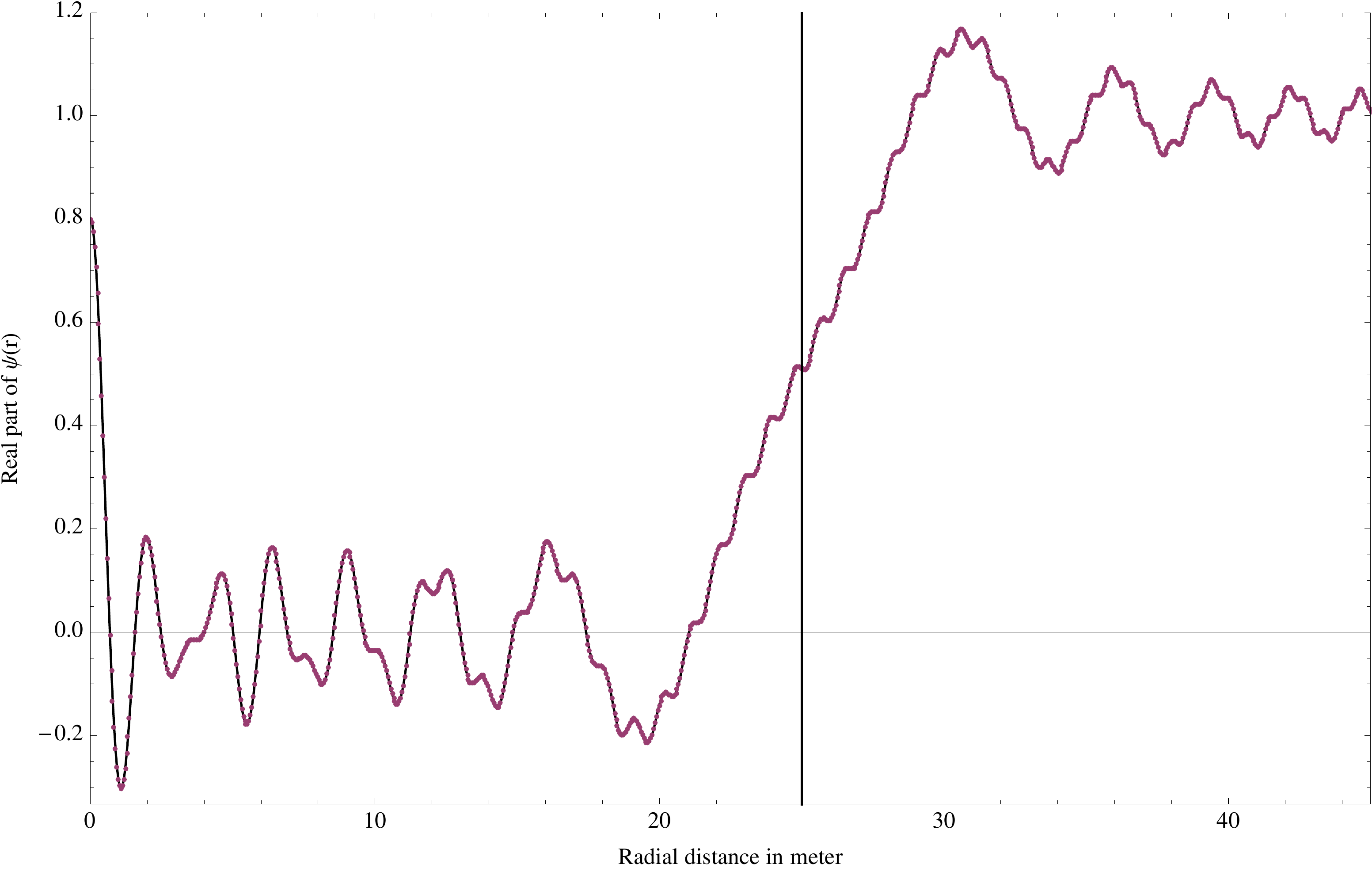} \\
\includegraphics[width=0.8\columnwidth]{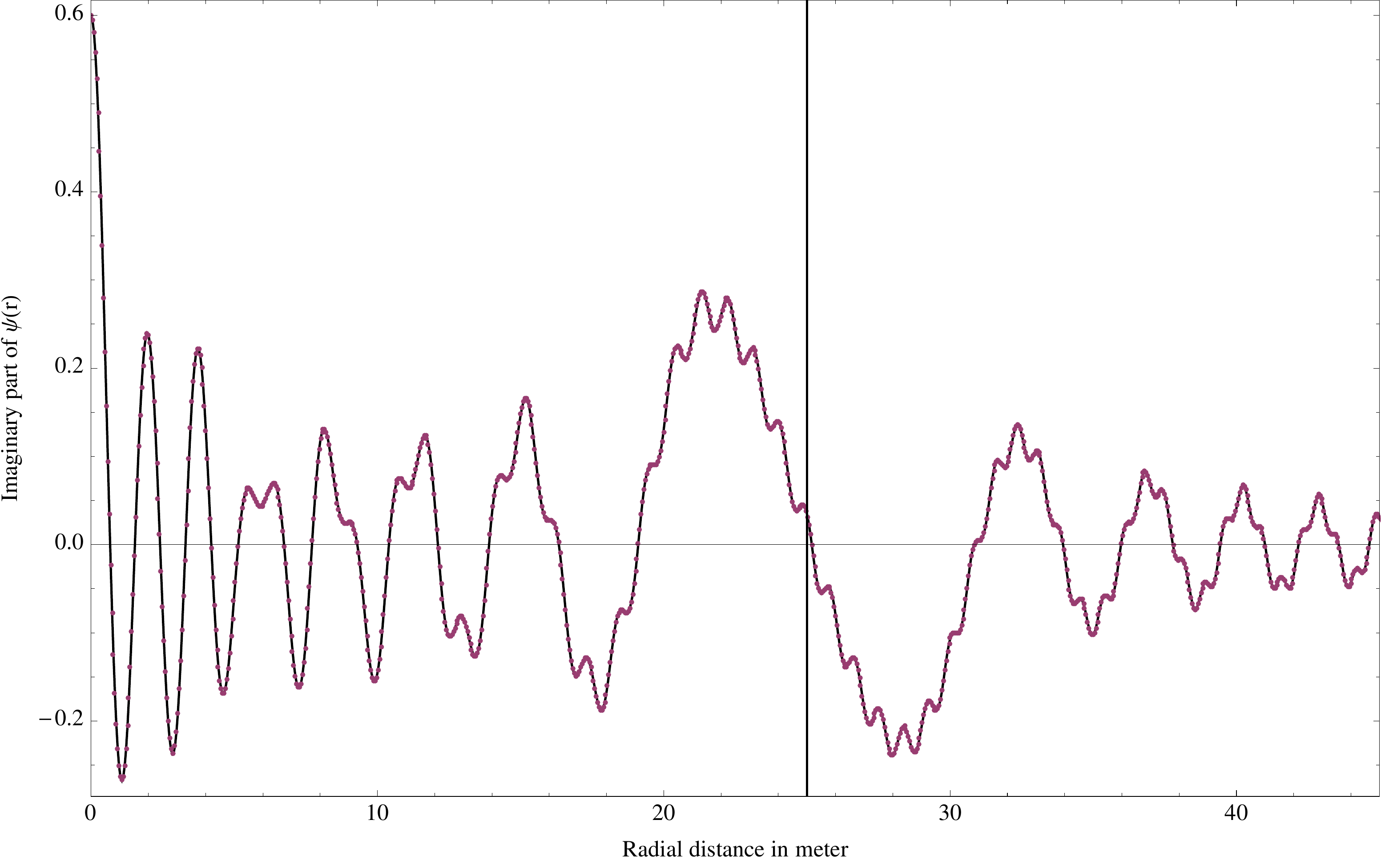} \\
\includegraphics[width=0.8\columnwidth]{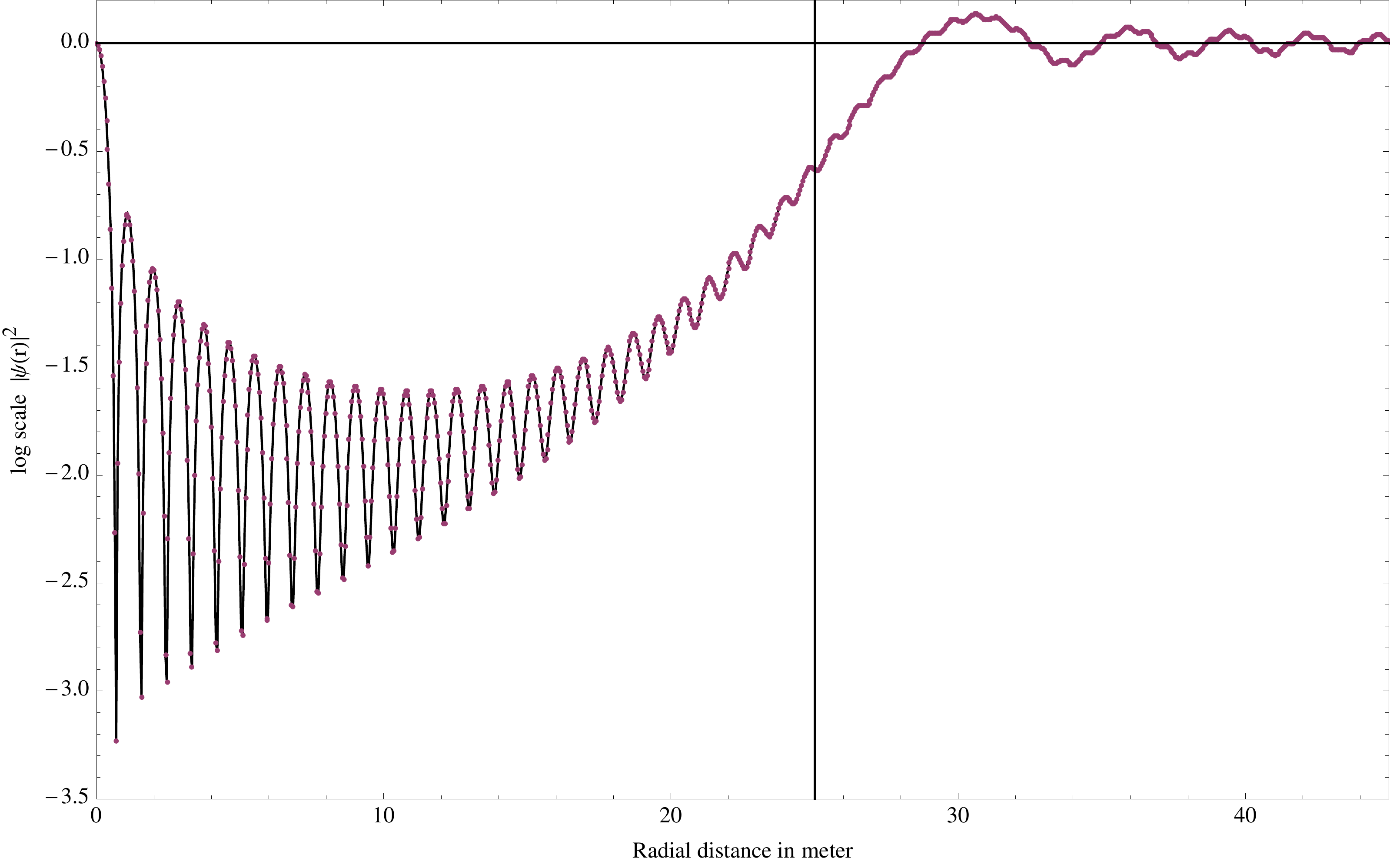} \\
\includegraphics[width=0.8\columnwidth]{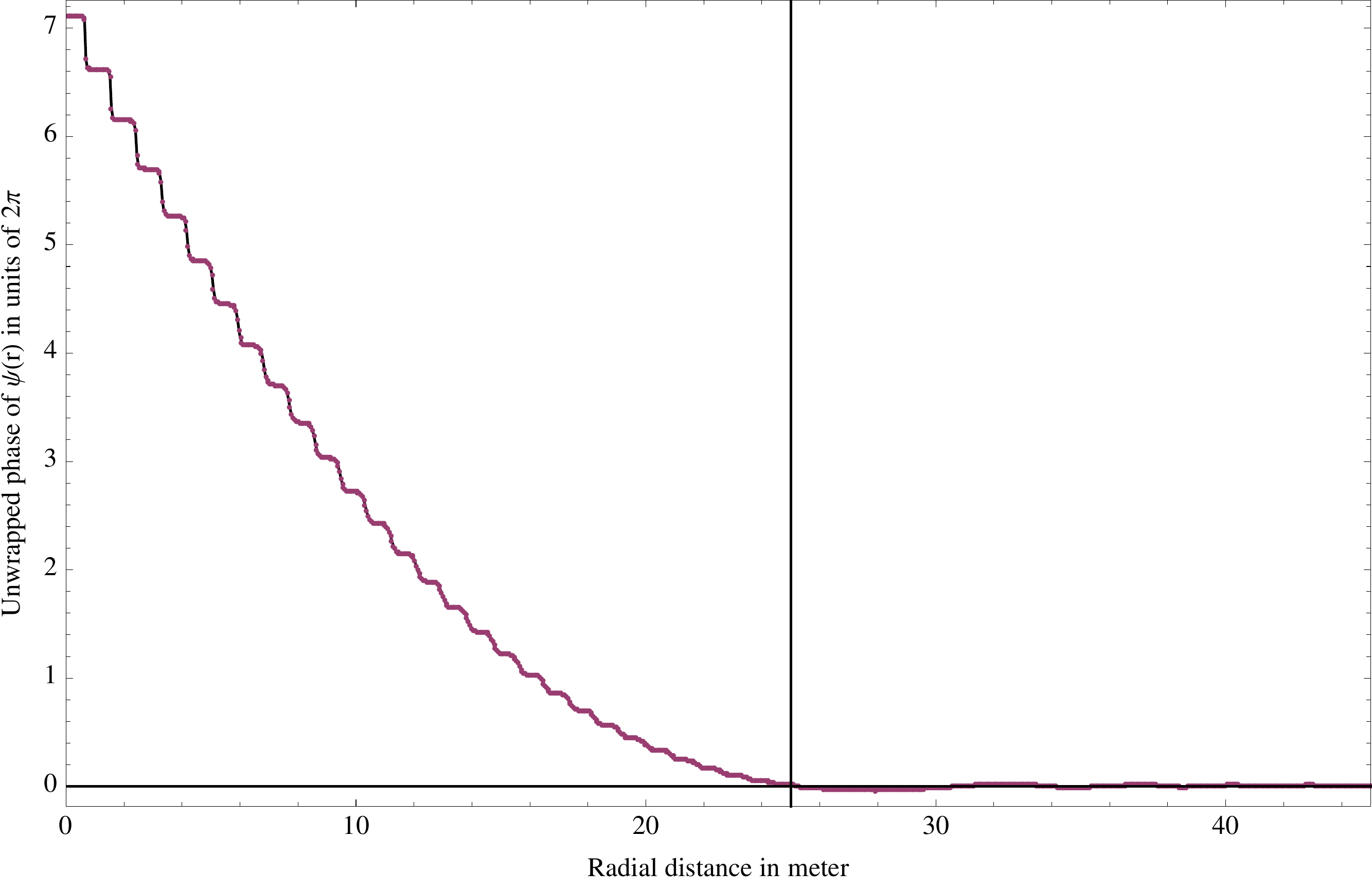} 
\end{tabular}
  \caption{ From top to bottom: real, imaginary, squared modulus and unwrapped phase of $\psi(r)$ for a circular occulter of radius $\Omega=25$ m (the vertical line) at $z=80 000$ km for $\lambda=0.55 \mu$ m. The continuous line represents the Lommel series, dots are 
  the result of a direct numerical computation of Eq. 1 using NIntegrate of \textit{Mathematica}}
  \label{precisionCalcul}
\end{figure}

 In Fig.\ref{precisionCalcul} we compare  the analytical and numerical
 results for an example of a circular occulter of 50 m diameter set at
 80 000 km, and for a wavelength of  550 nm. The Lommel series were
 computed for 100 terms, and there is no difference between the two
 methods. Results were drawn for the real, imaginary, modulus, and
 unwrapped phase of the Fresnel diffraction. Note that this is just an
 example, in particular, the real and imaginary part of the
 diffraction pattern are highly sensitive to a small variation of any
 of the parameters. The modulus of the intensity pattern clearly shows
 the visible Poisson spot at the center of the pattern. The unwrapped
 phase presents a strong phase variation in the center of the umbra
 and a  quieter structure outside the geometric umbra, that is for a
 region utilized by the telescope for the planet observation. The
 consistency between the two analytic and numerical results is
 recognized here  as a proof of the quality of the numerical
 calculation.

\section{Numerical calculation of linearly apodized basis functions $f_k(r)$}
\label{A2}

As   indicated in the body of the paper, we tried several forms for
the set of apodized basis functions  $f_k(r)$, and decided to use the
simple  trapezoidal functions described in the body of the paper
because they affect the final solution the least. We did not \rev{ succeed} in deriving  an analytic form for the Fresnel diffraction of these functions, but comforted  by  the results obtained on  the Fresnel diffraction of $\Pi(r/\Omega)$, we used the direct numerical calculation  of \textit{Mathematica}.

\begin{figure}
 \centering
\includegraphics[width=0.95\columnwidth]{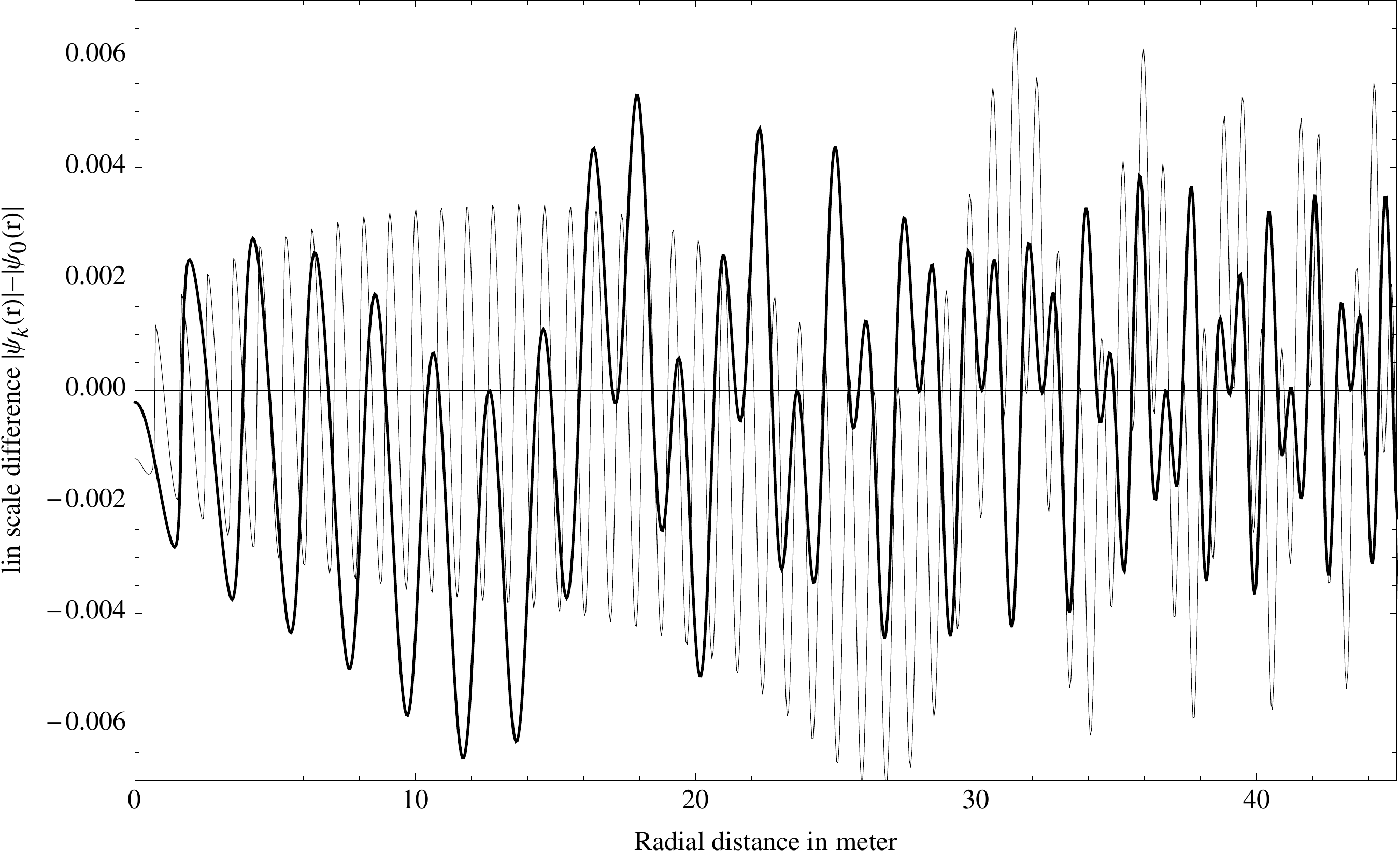} 
  \caption{ Differences of moduli of Fresnel diffractions for raw and linearly apodized functions  for  $\Omega$ values of 10 m (thick curve) and 10 m (thin curve).}
  \label{difference}
\end{figure}

\begin{figure}
 \centering
\includegraphics[width=0.95\columnwidth]{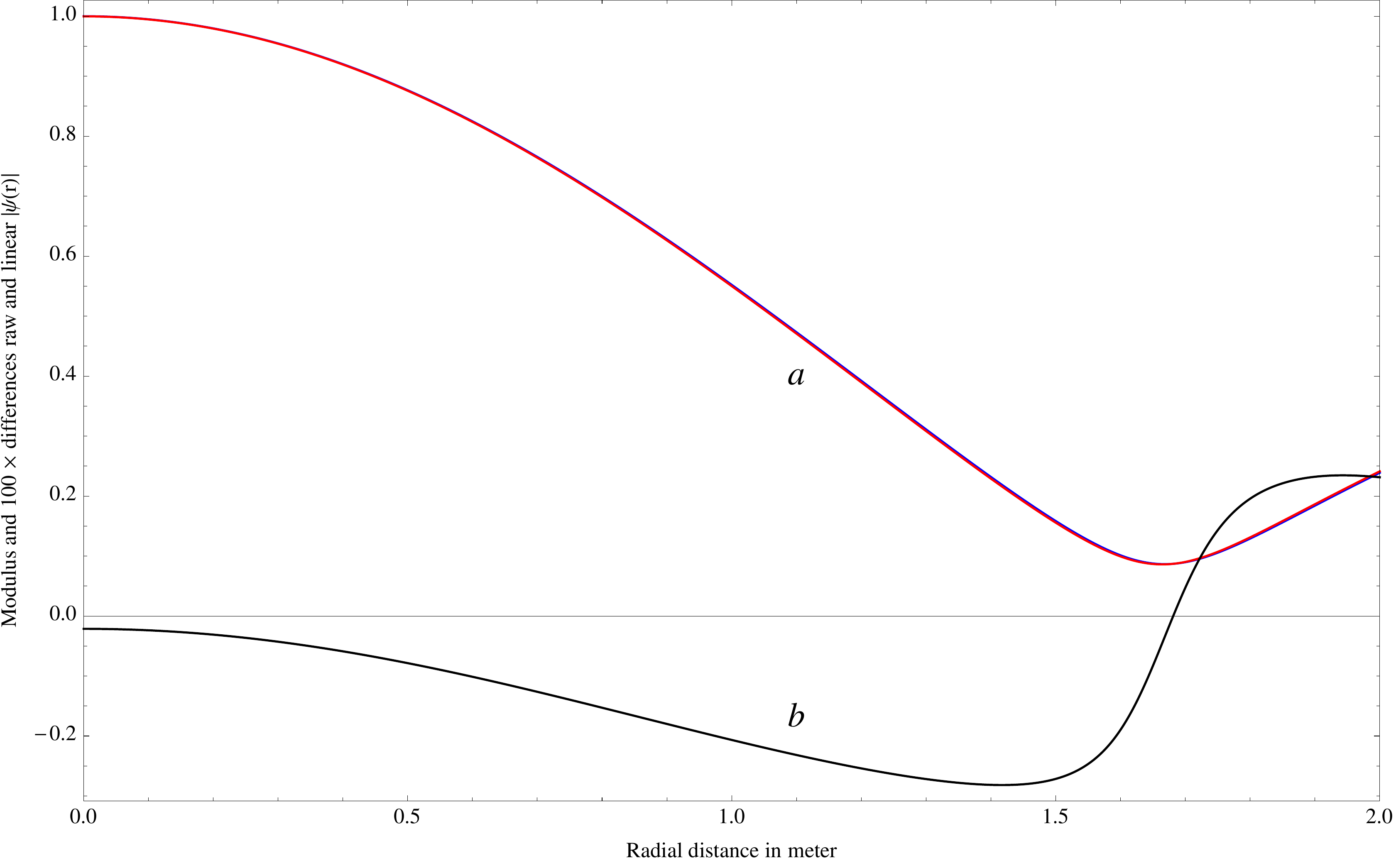} 
  \caption{  Representation for $r$ between 0 and 2 m of  the moduli of $\psi_k(r)$ of linearly apodized occulters of radii 10 m and 10.05 m (a). 
  The  two curves are almost over imposed, and curve (b) represents 100 times their differences.}
  \label{difference2}
\end{figure}

 We recall  that these functions are  linearly apodized disks of radii varying from 10 m to 25 m in steps of 5 cm, the apodization applies for  5 cm at the edge. The resulting $\psi_k(r)$ are very similar to those obtained for the functions   $\Pi(r/\Omega)$.  Differences between moduli are lower than $1 \%$, as shown in Fig. \ref{difference} for two extreme cases, but  essential to obtain proper results on  diffraction patterns. In Fig.  \ref{difference2}  we focus on the central part of the diffraction pattern, for $r \leq 2$ m. Curves are drawn for  a raw and linearly apodized occulter of 10 m and 10.05 m. 

\begin{acknowledgements} 
We want to thank the referee
Jeremy Kasdin for his very constructive comments, and in particular for 
his suggestion to study the effect of telescope misalignment.

This work
was granted access to the HPC and visualization resources of \emph{Centre
de Calcul Interactif} hosted by \emph{Université Nice Sophia
Antipolis} (\url{http://calculs.unice.fr/en}).
\end{acknowledgements}

\bibliographystyle{aa} 
 
\end{document}